\newcommand*{\ARXIV}{}
\newcommand*{\JQAS}{}

\ifdefined\JSA
\documentclass[]{interact}   
\fi
\ifdefined\JQAS
\documentclass[letterpaper,12pt]{article}   
\usepackage[authoryear,comma,longnamesfirst,sectionbib]{natbib}
\fi

\usepackage{comment}
\usepackage{centernot}

\newcommand{\CFilesBib}{Common.Files.Bib}

\usepackage{amsmath,amssymb,amscd,latexsym,dsfont,wasysym,bbm}
\usepackage{float}
\usepackage{color}
\usepackage{graphicx,subfigure}
\usepackage{multirow,multicol} 
\usepackage{verbatim}
\usepackage{psfrag}
\usepackage{comment}
\usepackage{makeidx}
\usepackage[]{algorithm}
\usepackage{algorithmicx}
\usepackage{algpseudocode}
\usepackage{cases}
 \usepackage{setspace}
\usepackage{pgfplots} 

\newcommand{\mr}[1]{\mathrm{#1}}
\newcommand{\tnr}[1]{{\textnormal{#1}}}

\newcommand{\mc}[1]{\mathcal{#1}}
\newcommand{\mf}[1]{\mathsf{#1}}
 
\newcommand{\ms}[1]{\mathds{#1}}

\newcommand{\ov}[1]{\overline{#1}}

\newcommand{\be}{\boldsymbol{e}}

\newcommand{\bx}{\boldsymbol{x}}

\newcommand{\balpha}{\boldsymbol{\alpha}}

\newcommand{\bdelta}{\boldsymbol{\delta}}

\newcommand{\btheta}{\boldsymbol{\theta}}

\newcommand{\brho}{\boldsymbol{\rho}}

\newcommand{\ie}{i.e.,~} 		
\newcommand{\eg}{e.g.,~}	

\newcommand{\argmax}{\mathop{\mr{argmax}}}

\newcommand{\set}[1]{\{#1\}}
\newcommand{\SET}[1]{\left\{#1\right\}}

\newcommand{\cd}{\cdot}
\newcommand{\ld}{\ldots}

\newcommand{\e}{\mr{e}}

\newcommand{\nchoosek}[2]{{{#1} \choose {#2}}}

\newcommand{\PR}[1]{\Pr\SET{#1}}       	
\newcommand{\pdf}{f}            			

\newcommand{\Ex}{\ms{E}}     			
\newcommand{\T}{^{\mf{T}}}            		
\newcommand{\dd}{\,\mr{d}}             		

\newcommand{\mcL}{\mc{L}}

\newcommand{\mcN}{\mc{N}}

\newcommand{\mcT}{\mc{T}}

\newcommand{\mcY}{\mc{Y}}

\newcommand{\mfP}{\mf{P}}

\newcommand{\Real}{\mathbb{R}}		

\ifdefined\JQAS
\usepackage{dgjournal}   
\fi

\pgfplotsset{compat=1.12}
\usetikzlibrary{positioning,shadows,shapes,fit,arrows,backgrounds}

\tikzstyle{rect_my} = [draw, rectangle, minimum width=2cm, text width=1.8cm, fill=gray!15, 
  text centered,  minimum height=.9cm]
\tikzstyle{square_my} = [draw, rectangle, minimum width=1cm, text width=0.8cm, fill=gray!15, 
  text centered,  minimum height=.9cm]
\tikzstyle{square_my_graph} = [draw, rectangle, minimum width=1.2cm, text width=1cm, fill=gray!15, 
  text centered,  minimum height=1.2cm]
\tikzstyle{circle_my} = [draw, circle, minimum width=1cm, text width=0.8cm, fill=gray!15, 
  text centered,  minimum height=.9cm]
\tikzstyle{circle_my_graph} = [draw, circle, minimum width=1.1cm, text width=.8cm, fill=gray!15, 
  text centered]
\tikzstyle{cloud_my} = [draw, shape=cloud, minimum width=1cm, text width=0.8cm, fill=gray!15, 
  text centered,  minimum height=1cm]

\tikzstyle{point_my} = [draw=none, minimum width=0cm, text width=0cm, fill=none, 
  text centered,  minimum height=0cm]    
\tikzstyle{line_my} = [draw, -latex]    
\tikzstyle{box_my}=[draw, minimum size=2em, text width=4.5em, text centered]
\tikzstyle{bigbox_my}=[draw, inner sep=15pt]
\tikzstyle{arrow_my} = [thick,->,>=stealth]
\tikzstyle{noarrow_my} = [thick,-,=>stealth]

\usepackage[title]{appendix}
\usepackage{booktabs}

\ifdefined\JSA      
\usepackage[natbibapa,nodoi]{apacite}
\fi

\usepackage[acronym,nonumberlist]{glossaries} 
\usepackage{hyperref}
\usepackage{empheq}
\usepackage{verbatim}
\usepackage{graphicx}
\usepackage{enumitem}
\newacronym[\glsshortpluralkey=PDFs,\glslongpluralkey=probability density functions]{pdf}{PDF}{probability density function}
\newacronym[\glsshortpluralkey=CDFs,\glslongpluralkey=cumulative density functions]{cdf}{CDF}{cumulative density function}
\newacronym[\glsshortpluralkey=CCDFs,\glslongpluralkey=complementary cumulative density functions]{ccdf}{CDF}{complementary cumulative density function}
\newacronym[\glsshortpluralkey=PMFs,\glslongpluralkey=probability mass functions]{pmf}{PMF}{probability mass function}
\newacronym[]{lhs}{l.h.s.}{left-hand side}
\newacronym[]{rhs}{r.h.s.}{right-hand side} 

\newacronym[]{bicm}{BICM}{bit-interleaved coded modulation}
\newacronym[]{bicmid}{BICM-ID}{BICM with iterative demapping}
\newacronym[]{cm}{CM}{coded modulation}
\newacronym[]{tcm}{TCM}{trellis-coded modulation}
\newacronym[]{mlc}{MLC}{multi-level coding}
\newacronym[]{pam}{PAM}{pulse amplitude modulation}
\newacronym[]{bpsk}{BPSK}{binary phase shift keying}
\newacronym[]{qam}{QAM}{quadrature amplitude modulation}
\newacronym[]{16qam}{16-QAM}{16-points quadrature amplitude modulation}
\newacronym[]{psk}{PSK}{phase shift keying}
\newacronym[\glsshortpluralkey=LLRs,\glslongpluralkey=logarithmic likelihood ratios]{llr}{LLR}{logarithmic likelihood ratio}
\newacronym[]{oc}{OC}{operating characteristic}

\newacronym[\glsshortpluralkey=MIs,\glslongpluralkey=mutual informations]{mi}{MI}{mutual information}
\newacronym[\glsshortpluralkey=GMIs,\glslongpluralkey=generalized mutual informations]{gmi}{GMI}{generalized mutual information}
\newacronym[]{eesm}{EESM}{exponential effective-SNR-mapping}
\newacronym[]{bicm-gmi}{BICM-GMI}{BICM generalized mutual information}
\newacronym[]{awgn}{AWGN}{additive white Gaussian noise}
\newacronym[]{bsc}{BSC}{binary symetric channel}
\newacronym[]{amc}{AMC}{adaptive modulation and coding}
\newacronym[]{csi}{CSI}{channel state information}
\newacronym[]{cqi}{CQI}{channel quality indicator}
\newacronym[]{kl}{KL}{Kullback-Leibler}
\newacronym[]{cmm}{CMM}{circular moment matching}
\newacronym[]{ga}{GA}{Gaussian approximation}

\newacronym[]{sp}{SP}{set-partitioning}
\newacronym[]{gsm}{GSM}{global system for mobile communications}
\newacronym[]{edge}{EDGE}{enhanced data rates for GSM evolution}
\newacronym[]{3gpp}{3GPP}{3rd generation partnership project}
\newacronym[]{umts}{UMTS}{Universal Mobile Telecommunication System}
\newacronym[]{lte}{LTE}{Long Term Evolution}
\newacronym[]{dvb}{DVB}{digital video broadcasting}
\newacronym[]{fdd}{FDD}{Frequency Division Duplexing}

\newacronym[\glsshortpluralkey=CCs,\glslongpluralkey=convolutional codes]{cc}{CC}{convolutional code}
\newacronym[\glsshortpluralkey=PCCCs,\glslongpluralkey=parallel concatenated convolutional codes]{pccc}{PCCC}{parallel concatenated convolutional code}
\newacronym[\glsshortpluralkey=TCs,\glslongpluralkey=turbo codes]{tc}{TC}{turbo code}
\newacronym{ldpc}{LDPC}{low-density parity-check}
\newacronym[]{ofdm}{OFDM}{orthogonal frequency-division multiplexing}
\newacronym[]{bep}{BEP}{bit-error probability}
\newacronym[]{wep}{WEP}{word-error probability}
\newacronym[]{sep}{SEP}{symbol-error probability}
\newacronym[]{pep}{PEP}{pairwise-error probability}
\newacronym[]{ttcm}{TTCM}{turbo-trellis coded modulation}
\newacronym[]{uep}{UEP}{unequal error protection}
\newacronym[\glsshortpluralkey=CENCs,\glslongpluralkey=convolutional encoders]{cenc}{CENC}{convolutional encoder}
\newacronym[]{mimo}{MIMO}{multiple-input multiple-output}
\newacronym[\glsshortpluralkey=SNRs,\glslongpluralkey=signal-to-noise ratios]{snr}{SNR}{signal-to-noise ratio}
\newacronym[\glsshortpluralkey=SINRs,\glslongpluralkey=the signal-to-interference-plus-noise ratios]{sinr}{SINR}{the signal-to-interference-plus-noise ratio}
\newacronym[]{msb}{MSB}{most-significative bit}
\newacronym[]{bcjr}{BCJR}{Bahl--Cocke--Jelinek--Raviv}
\newacronym[]{cbc}{CBC}{Colavolpe--Barbieri--Caire}
\newacronym[]{skr}{SKR}{Shayovitz--Kreimer--Raphaeli}
\newacronym[\glsshortpluralkey=SEDs,\glslongpluralkey=squared Euclidean distances]{sed}{SED}{squared Euclidean distance}
\newacronym[\glsshortpluralkey=EDs,\glslongpluralkey=Euclidean distances]{ed}{ED}{Euclidean distance}
\newacronym[\glsshortpluralkey=MEDs,\glslongpluralkey=minimum Euclidean distances]{med}{MED}{minimum Euclidean distance}
\newacronym[]{core}{CoRe}{constellation rearrangement}
\newacronym[]{pdl}{PDL}{parallel decoding of the individual levels}
\newacronym[\glsshortpluralkey=GCs,\glslongpluralkey=Gray codes]{gc}{GC}{Gray code}
\newacronym[]{brgc}{BRGC}{binary-reflected Gray code}
\newacronym[]{nbc}{NBC}{natural binary code}
\newacronym[]{fbc}{FBC}{folded-binary code}
\newacronym[]{bsgc}{BSGC}{binary semi-Gray code}
\newacronym[]{msp}{MSP}{modified set-partitioning}
\newacronym[]{ssp}{SSP}{semi set-partitioning}
\newacronym[]{fhd}{FHD}{free Hamming distance}
\newacronym[]{mfhd}{MFHD}{maximum free Hamming distance}
\newacronym[]{ods}{ODS}{optimal distance spectrum}
\newacronym[]{iud}{i.u.d.}{independent and uniformly distributed}
\newacronym[]{ud}{u.d.}{uniformly distributed}
\newacronym[]{iid}{i.i.d.}{independent, identically distributed}
\newacronym[]{ami}{AMI}{accumulated mutual information}
\newacronym[]{bico}{BICO}{binary-input continuous-output}
\newacronym[]{gh}{GH}{Gauss--Hermite}
\newacronym[]{gum}{GUM}{Gaussian--uniform mixture}

\newacronym[\glsshortpluralkey=BSs,\glslongpluralkey=base-stations]{bs}{BS}{base-station}
\newacronym[\glsshortpluralkey=MSs,\glslongpluralkey=mobile-stations]{ms}{MS}{mobile-stations}

\newacronym[]{phy}{PHY}{physical layer} 
\newacronym[]{rlc}{RLC}{Radio-Link control} 
\newacronym[]{ran}{RAN}{Radio Access Network} 
\newacronym[]{llc}{LLC}{logical link control} 
\newacronym[]{tcp}{TCP}{transmission control protocol} 
\newacronym[]{mac}{MAC}{media access control} 
\newacronym[]{fft}{FFT}{fast Fourier transform} 
\newacronym[]{ft}{FT}{Fourrier transform}
\newacronym[]{cf}{CF}{characteristic function} 
\newacronym[]{mgf}{MGF}{moment generating function} 
\newacronym[]{ee}{EE}{energy efficiency} 
\newacronym[]{eb}{EB}{energy per bit}
\newacronym[]{kkt}{KKT}{Karush--Kuhn--Tucker} 
\newacronym[]{mcs}{MCS}{modulation/coding scheme} 
\newacronym[]{fec}{FEC}{forward error correction}
\newacronym[]{arq}{ARQ}{automatic repeat request}
\newacronym[]{harq}{HARQ}{hybrid ARQ}
\newacronym[]{tarq}{TARQ}{truncated HARQ}
\newacronym[]{ir}{IR}{incremental redundancy}
\newacronym[]{rpr}{RR}{repetition redundancy}
\newacronym[]{rrharq}{RR-HARQ}{repetition redundancy HARQ}
\newacronym[]{irharq}{IR-HARQ}{incremental redundancy HARQ}
\newacronym[]{ack}{ACK}{positive acknowledgment}
\newacronym[]{nack}{NACK}{negative acknowledgment}
\newacronym[]{hol}{HoL}{head of the line}
\newacronym[]{crc}{CRC}{cyclic redundancy check}
\newacronym[]{dp}{DP}{dynamic programming}
\newacronym[]{gp}{GP}{geometric programming}
\newacronym[]{per}{PER}{packet error rate}
\newacronym[]{ber}{BER}{bit error rate}
\newacronym[]{op}{OP}{outage probability}
\newacronym[]{spa}{SPA}{saddle-point approximation}
\newacronym[]{mrc}{MRC}{maximum ratio combining}
\newacronym[]{mdp}{MDP}{Markov decision process}
\newacronym[]{lp}{LP}{linear programming}
\newacronym[]{pomdp}{POMDP}{partially observable Markov decision process}
\newacronym[]{psimdp}{PSI-MDP}{partial state information Markov decision process}
\newacronym[]{scpp}{SCPP}{stochastic shortest path problem}

\newacronym[]{forw}{frwd}{forward}
\newacronym[]{feed}{fdbk}{feedback}

\newacronym[]{mm}{MM-HARQ}{multi-message HARQ}
\newacronym[]{xp}{XP-HARQ}{cross-packet HARQ}
\newacronym[]{ts}{TS}{time-sharing}
\newacronym[]{sc}{SC}{superposition coding}
\newacronym[]{sbrq}{SBRQ}{systematic backward retransmission}
\newacronym[]{brq}{BRQ}{backward retransmission}
\newacronym[]{lharq}{L-HARQ}{layer-coded HARQ}
\newacronym[]{anlharq}{AoN-HARQ}{all-or-none L-HARQ}
\newacronym[]{vlharq}{VL-HARQ}{variable-length HARQ}

\newacronym[]{pp}{PPP}{point process}
\newacronym[]{ppp}{PPP}{Poisson point process}

\newacronym[]{fide}{FIDE}{F\'ed\'eration Internationale des \'Echecs}
\newacronym[]{fifa}{FIFA}{F\'ed\'eration Internationale de Football Association}
\newacronym[]{fivb}{FIVB}{F\'ed\'eration Internationale de Volleyball}
\newacronym[]{epl}{EPL}{English Premier League}
\newacronym[]{nhl}{NHL}{National Hockey League}
\newacronym[]{shl}{SHL}{Swedish Hockey League}
\newacronym[]{nfl}{NFL}{National Football League}
\newacronym[]{ipl}{IPL}{Indian Premier League}
\newacronym[]{nba}{NBA}{National Basketball Association}
\newacronym[]{mls}{MLS}{Major League Soccer}

\newacronym[]{sg}{SG}{stochastic gradient}
\newacronym[]{lms}{LMS}{least mean squares}
\newacronym[]{rls}{RLS}{recursive least squares}
\newacronym[]{vss}{VSS}{variable step-size}
\newacronym[]{hfa}{HFA}{home-field advantage}
\newacronym[]{ha}{HA}{home advantage}
\newacronym[]{mov}{MOV}{margin of victory}
\newacronym[]{ac}{AC}{Adjacent Categories}
\newacronym[]{cl}{CL}{Cumulative Link}
\newacronym[]{glm}{GLM}{Generalized Linear Models}
\newacronym[]{nn}{NN}{Neural Networks}
\newacronym[]{rps}{RPS}{Ranked Probability Score}
\newacronym[]{mse}{MSE}{Mean Squared Error}
\newacronym[]{mmse}{MMSE}{Minimum Mean Squared Error}
\newacronym[]{rmse}{RMSE}{Root Mean Squared Error}
\newacronym[]{map}{MAP}{maximum a posteriori}
\newacronym[]{ml}{ML}{maximum likelihood}
\newacronym[]{loo}{LOO}{leave-one-out}
\newacronym[]{alo}{ALO}{approximate leave-one-out}
\newacronym[]{logo}{LOGO}{leave-one-game-out}
\newacronym[]{alogo}{ALOGO}{approximate leave-one-game-out}
\newacronym[]{msd}{MSD}{mean-square deviation}
\newacronym[]{lop}{LOP}{linear ordering problem}
\newacronym[]{so}{SO}{shootouts}
\newacronym[]{rt}{RT}{regulation time}
\newacronym[]{ot}{OT}{overtime}
\newacronym[]{rr}{RR}{round-robin}
\newacronym[]{irt}{IRT}{item-response theory}

\newacronym[]{dmp}{DMP}{discretized message passing}
\newacronym[]{mp}{MP}{message passing}
\newacronym[]{ep}{EP}{expectation propagation}
\newacronym[]{em}{EM}{expectation maximization}
\newacronym[]{hmm}{HMM}{hiden Markov models}

\newacronym[]{svd}{SVD}{singular values decomposition}

\newacronym[]{skf}{SKF}{Simplified Kalman Filter}
\newacronym[]{vskf}{vSKF}{\emph{vector-covariance} Simplified Kalman Filter}
\newacronym[]{sskf}{sSKF}{\emph{scalar-covariance} Simplified Kalman Filter}
\newacronym[]{fskf}{fSKF}{\emph{fixed-variance} Simplified Kalman Filter}
\newacronym[]{kf}{KF}{Kalman Filter}
\newacronym[]{gelo}{G-Elo}{Generalized Elo}
\newacronym[]{mvdr}{MVDR}{Minimum Variance Distortionless Response}
\newacronym[]{music}{MUSIC}{Multiple Signal Classification}
\newacronym[]{cp}{CP}{Canonical Polyadic}

\newacronym[]{tpb}{TPB}{tensor-product-basis}

\newtheorem{proposition}{Proposition}

\newtheorem{example}{Example}

\title{New insights into Elo algorithm\\ for practitioners and statisticians}
\author{Leszek Szczecinski}

\begin{document}

\ifdefined\JSA
\maketitle    
\fi
\ifdefined\JQAS
\originalmaketitle  
\fi

\begin{abstract}
This work reconciles two perspectives on the Elo ranking that coexist in the literature: the practitioner's view as a heuristic feedback rule, and the statistician's view as online maximum likelihood estimation via stochastic gradient ascent. Both perspectives coincide exactly in the binary case (iff the expected score is the logistic function). However, estimation noise forces a principled decoupling between the model used for ranking and the model used for prediction: the effective scale and home-field advantage parameter must be adjusted to account for the noise. We provide both closed-form corrections and a data-driven identification procedure. For multilevel outcomes, an exact relationship exists when outcome scores are uniformly spaced, but approximations are preferred in general: they account for estimation noise and better fit the data.

The decoupled approach substantially outperforms the conventional one that reuses the ranking model for prediction, and serves as a diagnostic of convergence status. Applied to six years of FIFA men's ranking, we find that the ranking had not converged for the vast majority of national teams.

The paper is written in a semi-tutorial style accessible to practitioners, with all key results accompanied by closed-form expressions and numerical examples.
\end{abstract}

\newpage
\section{Introduction}\label{Sec:Introduction}

Ranking in sports is used to decide which teams should be promoted/relegated between leagues or how to fix the competition format; it also provides the quick understanding of the relative strengths of the teams/players. It is thus of fundamental importance and already attracted consistent interest in sports, \eg chess, football, e-sports as well as in academic circles~\cite{Glickman95, Aldous17,Lasek18,Morel25}.

Ranking algorithms/strategies were often devised by practitioners who, guided by their intuition and understanding of the competition, were able to propose practical solutions. Among the many algorithms, the Elo ranking~\cite{Elo78_Book} stands out due to its remarkable resilience and scope of application. Introduced by Arpad Elo in the context of chess~\cite{Elo78_Book} and today used by governing bodies of \gls{fide}~\cite{fide_calculator} and  \gls{fifa}~\cite{fifa_rating}, it has also been applied to analyze theoretically football~\cite{eloratings.net,Szczecinski22a,Csato24}, basketball, American football \cite{Silver20}, tennis \cite{Kovalchik20}, and beyond.

This unquestionable success is easy to understand: the Elo algorithm is simple to implement, transparent in its logic, and adapts naturally to settings where competitors' abilities evolve over time. However, we must agree with Aldous~\cite{Aldous17}, that Elo algorithm is indeed a case of  ``neglected topic in applied probability'' whose theoretical understanding deserve more attention. 

This is particularly true because the literature related to Elo ranking was dominated by the \emph{practitioner's} perspective, which sees the algorithm as a heuristic feedback rule: skills are updated in proportion to the gap between an observed match score and a pre-defined expected score \cite{Hvattum10,Lasek13,Cortez26}. No probabilistic model is required, and the expected score function can be chosen freely, as was done in the original formulation by A. Elo \cite{Elo78_Book}. This also invites modification which mix probabilistic models and heuristics \cite{Kovalchik20}, \cite{Ingram21}. 

Although heuristics may be useful, they often lack the proper methodology to evaluate \emph{objectively} the ranking as this usually requires some kind of predictive capability, which, in turn needs a probabilistic model linking the skills with the outcomes.

Recognizing this fundamental weakness, the \emph{statistician's} perspective,\footnote{We use the term ``statistician" rather loosely to denote any mathematically-oriented appraoch.} always starts with the probabilistic model linking unobservable skills of competitors to matches outcomes, and the algorithm is interpreted as an online implementation of \gls{ml} estimation via \gls{sg} ascent~\cite{Kiraly17,Szczecinski20} or through more sophisticated Bayesian update \cite{Glickman99,Szczecinski23a,Ingram21}.

These two perspectives are not always distinguished carefully in the literature, and conflating them can lead to difficulties in interpreting and evaluating results. This is true for matches with two outcomes (win/loss) but even more so when multi-level outcomes must be used for ranking (\eg win/draw/loss).

It should be noted that, in a large body of work on the ranking and modelling in sports, the statistician's approach usually produces new algorithms, which may be similar to, but are not the same as the simple Elo algorithm. Needless to say, among the many proposals for ranking, \eg \cite{Rao67,Davidson77,Karlis08,Lasek20,Egidi20,Szczecinski20,Kovalchik20,Ingram21,Angelini21,Szczecinski22}, none succeeded in dethroning the Elo algorithm. Thus, following the principle ``if you cannot beat them, join them'', the objective of our work is to reconcile the two perspectives with two main goals: a) provide practitioners with tools to understand and interpret the Elo ranking, and b) give statisticians a less rigid framework to analyze the Elo algorithm by decoupling the ranking algorithm from the predictive model.

To bridge the two perspectives in a constructive way, we focus on what is practically useful, build on prior work on the statistical foundations of Elo~\cite{Kiraly17,Aldous17,Szczecinski20,Szczecinski22,Zanco24} and on the evaluation of Elo-based rankings~\cite{Szczecinski22a}, and make the following contributions:
\begin{enumerate}
    \item We show a self-contained derivation of the equivalence between the practitioner's Elo algorithm and \gls{ml} estimation, for both binary and multilevel outcomes (Sec.~\ref{Sec:practitioner.meets.statistician} and Sec.~\ref{Sec:ordinal.models}).
    

    \item We quantify in simple terms the dependence of the estimation noise and the convergence speed on the adaptation step $K$ and the scale $s$, generalizing results of \cite{Zanco24}.

    \item We quantify the effect of estimation noise on predictive performance and show how the effective scale and the \gls{hfa} parameter must be adjusted to account for it (Sec.~\ref{Sec:Performance evaluation}). The resulting correction clarifies and extends results in~\cite{Ingram21,Sonas11} by making explicit the role of the skills' dispersion.

    \item We postulate a \emph{model decoupling} principle (Sec.~\ref{Sec:Performance evaluation} and Sec.~\ref{Sec:model.decoupling}) linking the Elo algorithm to the \gls{ac} model that can used for prediction. 
    
    We demonstrate via synthetic and real FIFA data that simple closed-form formulas defining the \gls{ac} model already capture most of the prediction gain.
\end{enumerate}

The manuscript is structured in a semi-tutorial fashion, to be accessible to practitioners.
Section~\ref{Sec:Elo.algorithm} treats the binary-outcome case, establishing the convergence properties, provides probabilistic interpretation of skills, and introduces the idea of scale adjustment due to estimation noise. Section~\ref{Sec:Multilevel.outcomes} extends the framework to multilevel outcomes, identifies the \gls{ac} model underlying the Elo algorithm, and develops the model-decoupling approach with examples on synthetic and \gls{fifa} data. Numerical examples are integrated into text and, to simplify the flow, some mathematical derivations are relegated to Appendices.

\section{Ranking from pairwise comparisons}\label{Sec:Elo.algorithm}

We consider a scenario in which $M$ players/teams indexed with $m=1, \ld, M$ face each other in one-on-one matches indexed with $t=1,\ld,T$, where $T$ is the total number of matches. The results of the matches are denoted as $y_t\in\mcY, t=1,\ld,T$, where $\mcY=\set{o_0,o_1, \ld, o_{L-1}}$ is the set of possible match outcomes which are ordered according to their importance, \ie $o_l$ is more important than $o_{l-1}$. Although the outcomes are not numerical, we can map them into their indices, \ie use $\mcY=\set{0,\ld, L-1}$, which yields a convenient notation applied henceforth. 

The objective of the ranking is, after observing the matches $y_t$, to characterize the performance of each player using a scalar parameter called ``skills'' (also known as ``merit'' or ``strength'' \cite{Langeville12_book}, \cite{Newman23}), which is denoted by $\theta_{t,m}\in\Real$.

By sorting the skills, we obtain a ranking (\ie we take for granted that $\theta_{t,m}>\theta_{t,n}$ means that the player $m$ is ``better'' than the player $n$; we will provide a more rigorous interpretation in Sec.~\ref{Sec:Performance evaluation}). The dependence of the skills on the index of the match $t$, may be interpreted as the time variability of skills (\eg due to improvement after training or degradation after injuries).

Using $i_t$ and $j_t$ to identify, respectively, home and away players, the most popular assumption is that the difference 
\begin{align}
\label{z.t.define}
    z_t 
    &= \theta_{t,i_t}-\theta_{t,j_t}
\end{align}
suffices to model/predict the match result. We do not consider other models such as those shown in \cite{Glickman25} which, in addition to the strengths' difference, $z_t$, also take into account the (average) strengths of the players. 

Since the difference between the skills $\theta_{t,m}-\theta_{t,n}$ tells us ``how much'', at time $t$ the player $m$ is better than the player $n$, this approach is also known as ``power-ranking''. 

We assume that, by increasing the value of $y_t$, we increase the importance of the outcome from the point of view of the home player, $i_t$, and vice versa, decrease its importance from the point of view of the away player $j_t$. For example, the win of the home player $i_t$ is equivalent to the loss of the away player $j_t$. The change of the perspective may be done using $(\cd)'$ to denote variable after swapping the home and away players, \ie $i'_t=j_t$, $j'_t=i_t$, $z'_t=-z_t$, and $y'_t=L-1-y_t$. 

We gather skills in a column vector $\btheta_t=[\theta_{t,1},\ld,\theta_{t,M}]\T$, where $(\cd)\T$ denotes transpose, and use a scheduling vector $\bx_t=\be_{i_t}-\be_{j_t}$, where $\be_{i}$ is the $i$-th canonical basis vector. This simplifies the notation of skills subtractions:
\begin{align}
\label{z.t.x.tdefine}
    z_t 
    &=\bx\T_t \btheta_t.
\end{align}

The objective of the on-line ranking algorithms is to estimate $\btheta_t$ after each match $t$, taking into account all previous outcomes, \ie
\begin{align}
\label{hat.btheta.t}
   \btheta_{t+1} &= \phi(y_{t}, y_{t-1}, y_{t-2}, \ld),
\end{align}
however, in practice, we want to use simple algorithms, and the most popular on-line rankings implement \eqref{hat.btheta.t} recursively
\begin{align}\label{hat.theta.recursive}
    \btheta_{t+1} &= \phi(\btheta_{t}, y_t),
\end{align}
that is, $\btheta_{t}$ -- the estimate of the skills before the match $t$, and the outcome of the match $y_t$ are used to obtain the updated estimate of the skills after the match $t$.

\subsection{Elo ranking in binary matches}\label{Sec:ranking.binary}

Let us assume that we deal with binary matches, \ie $L=2$, where $y_t=0$ denotes the loss of the home player $i_t$ and $y_t=1$ indicates their win. Later, we will deal with the case of $L>2$, but the binary case simplifies the discussion and allows us to easily understand two perspectives on the well-known Elo ranking.

\subsubsection{Practitioner's perspective}\label{Sec:Practitioner.perspective}

To obtain the Elo ranking, the practitioner needs to take the following steps:
\begin{enumerate}
    \item
    Assign a numerical \emph{score} $\rho_y$ to each match's outcome $y$,
    \begin{align}
        y \mapsto \rho_y\in\set{0,1},\quad y\in\mcY,
    \end{align} 
    where the highest importance $y$ has a higher score, that is, $\rho_y$ increases monotonically with $y$. In the binary case, we thus may use
    \begin{align}
    \label{rho.y_t.binary}
        \rho_{y_t}=y_t,
    \end{align} 
    \ie $\rho_0=0$ and $\rho_1=1$.
    
    \item 
    Define the \emph{expected score} 
    \begin{align}
        G(z_t) \in[0,1],
    \end{align}
    where we require $G(z_t)$ to increase monotonically with $z_t$ because we ``expect'' larger score when the skills' difference grows; in the limit we have the following:
    \begin{align}
        \lim_{z\rightarrow\infty} G(z)&=1,\\
        \lim_{z\rightarrow-\infty} G(z)&=0.
    \end{align}

    Moreover, the expected scores of the home player and the away players perspectives must add to one, 
    \begin{align}
    \label{G(z)_G(z')=1}
        G(z_t)+G(z'_t)&=1,
    \end{align}
    and since $z'_t=-z_t$, the expected score is ``antisymmetric''
    \begin{align}
    \label{G(z).is.antisymmetric}
        G(z_t)&=1-G(-z_t).
    \end{align}
    
    \item Use the \emph{positive feedback}, where the skills are increased proportionally to the gap between the observed and expected scores:
    \begin{align}
    \theta_{t+1,i_t} 
    &= \theta_{t,i_t} + \mu \big[\rho_{y_t} - G(z_t) \big]\\
\label{Elo.practitioner.i_t}
    &= \theta_{t,i_t} + \mu \big[y_t - G(z_t) \big].
\end{align}
  That is, skills are increased if the gap is positive, are decreased if the gap is negative, and the parameter $\mu>0$, a step, controls the magnitude of the update.

  Similarly, from the point of view of the away-player we write
  \begin{align}
\label{Elo.practitioner.j_t}
    \theta_{t+1,j_t} 
    &= \theta_{t,j_t} + \mu \big[\rho_{y'_t} - G(z'_t) \big],\\
    &= \theta_{t,j_t} + \mu \big[1-y_t - G(-z_t) \big],
    \end{align}
    which, from \eqref{G(z).is.antisymmetric}, becomes
    \begin{align}
    \label{Elo.practitioner.j_t.2}
    \theta_{t+1,j_t} 
    &= \theta_{t,j_t} - \mu \big[y_t - G(z_t) \big],
    \end{align}
    and,
    using the scheduling vector $\bx_t$,  \eqref{Elo.practitioner.i_t} and \eqref{Elo.practitioner.j_t.2}
    may be compactly written as
    \begin{align}
    \label{Elo.practitioner.compact}
    \btheta_{t+1} = \btheta_t + \mu \bx_t\big[y_t -G(z_t) \big].
    \end{align}
\end{enumerate}

Treating $y_t$ as realizations of random variables $Y_t$ with distributions that depend on $z_t$, $G(z_t)$ is a \emph{mathematical expectation} of the score $\rho_{y_t}$
\begin{align}
\label{expectation.definition}
    G(z_t) 
    &= \Ex_{Y_t|z_t}[\rho_{Y_t}]= \sum_{y\in\mcY}\PR{Y_t=y|z_t}\rho_y,
\end{align}
which, for binary matches ($\rho_0=0$, $\rho_1=1$) produces
\begin{align}
\label{Ex=Pr.1}
    G(z_t)&=\PR{Y_t=1|z_t}.
\end{align}

Thus, in matches with binary outcomes, the expected score defines the probability of the win for the home player, conditioned on the difference in skills, $z_t$.

However, from the practitioner's point of view, the ``expected score'' $G(z_t)$ may be used informally without precise mathematical meaning, which allows the Elo algorithm to exist without defining the probabilistic models. In fact, this is what is usually done. For example, in the \gls{fide} \cite{fide_calculator} and \gls{fifa} rankings \cite{fifa_rating} $G(z_t)$ is called ``expected score'' without any mention of the model which is necessary to calculate the mathematical expectation.

\subsubsection{Statistician's perspective}\label{Sec:Statistician.perspective}

Using the practitioner's perspective in Sec.~\ref{Sec:Practitioner.perspective}, we obtain, in \eqref{Ex=Pr.1}, a probabilistic model linking the skills difference $z_t$ to the random match outcome, $Y_t$. However, this is a side effect that was not required to derive the algorithm.

On the other hand, the statistician will always start by defining the model 
\begin{align}
    \PR{Y_t=y|\btheta^*}&=\mfP_y(z^*_t),\quad y\in\set{0,1},
\end{align}
where $\btheta^*$ are unknown ``true'' skills we want to estimate and the probability depends on the skills's difference $z^*_t=\bx\T_t\btheta^*$.\footnote{A practical way of thinking about $\btheta^*$ is as result of estimation when the skills do not change in time and the number of matches between all players goes to infinity.} 

The practitioner's assumptions are reproduced: $\mfP_1(z^*_t)$ must increase monotonically in $z^*_t$ (home win become more probable when  $z^*_t$ grows), and home-away swap ($z^*_t\leftarrow -z^*_t$) does not change the probability ($y_t\leftarrow L-1-y_t$, \eg home win before swap becomes away win after swap)
\begin{align}
\label{P_y(z).symmetry}
    \mfP_y(z)=\mfP_{L-1-y}(-z),
\end{align}
which, in the binary case implies that
\begin{align}
\label{P_1(z)=P_0(-z)}
    \mfP_1(z) &= \mfP_0(-z)=1-\mfP_1(-z).
\end{align}

The skills, treated as unknown parameters of the model, can then be inferred using a preferred estimation method. For example, we may apply the \gls{ml} principle, 
\begin{align}
\label{ML.definition}
    \hat\btheta&=\argmax_{\btheta} \sum_{t\in\mcT} \ell_{y_t}(\bx\T_t\btheta)
\end{align}
where $\mcT$ is a batch of indices and
\begin{align}
    \ell_y(z_t)&= \log\mfP_{y}(z_t)
\end{align}
is the log-likelihood of $z_t=\bx_t\T\btheta$ for the observed match results $Y=y$.

If instead of a batch optimization  \eqref{ML.definition} we opt for an on-line approach \eqref{hat.theta.recursive}, we can apply the \gls{sg} principle to the \gls{ml} problem in  \eqref{ML.definition}. It boils down to the sequential update of the skills according to the following \gls{ml}+\gls{sg} algorithm:
\begin{align}
    \btheta_{t+1} 
    &= \btheta_{t} + \mu\nabla_{\btheta_t} \ell_{y_t}(z_t)\\
\label{ML.update}
    &= \btheta_{t} + \mu \bx_t \dot\ell_{y_t}(z_t),
\end{align}
where
\begin{align}
    \label{dot.ell.definition}
    \dot\ell_{y}(z_t)&=\frac{\dd}{\dd z} \ell_{y}(z)\big|_{z=z_t}
    = \frac{\dot\mfP_y(z_t)}{\mfP_y(z_t)},
\end{align}
and $\mu$ is the adaptation step.

Using \eqref{P_1(z)=P_0(-z)} we obtain
\begin{align}
\label{dotP_0(z)=-dotP_1(z_}
    \dot\mfP_0(z) &= -\dot\mfP_1(z),
\end{align}
which allows us to calculate \eqref{dot.ell.definition} as
\begin{align}
    \dot\ell_{y_t}(z_t) &= (1-y_t)\frac{\dot\mfP_0(z_t)}{\mfP_0(z_t)}
    +y_t \frac{\dot\mfP_1(z_t)}{\mfP_1(z_t)}\\
    &=(y_t-1)\frac{\dot\mfP_1(z_t)}{1-\mfP_1(z_t)}
    +y_t \frac{\dot\mfP_1(z_t)}{\mfP_1(z_t)}\\
\label{g_t.final}
    &=[y_t-\mfP_1(z_t)]\phi(z_t)\\
\label{xi.definition}
    \phi(z_t)&=\frac{\dot\mfP_1(z_t)}{[1-\mfP_1(z_t)]\mfP_1(z_t)}.
\end{align}
which used in \eqref{ML.update} yields
\begin{align}
    \btheta_{t+1} 
\label{ML.update.full}
    &= \btheta_{t} + \mu \bx_t [y_t-\mfP_1(z_t)]\phi(z_t).
\end{align}

Both practitioner and statistician know that, to guarantee the convergences of the algorithm we need a ``sufficiently small'' $\mu$ which is often determined heuristically (with values $\mu<0.1$ usually working ``well'' but the exact conditions for convergence may be more involved \cite{Zanco24}). On the other hand, only recognizing the algorithm as a \gls{sg} maximization, we know that, we also need a concave form of $\ell_y(z)$ -- a condition which is satisfied for $\mfP_1(z)=\mcL(z)$ and $\mfP_1(z)=\Phi(z)$.\footnote{These are two most popular cases used in ranking, but other log-concave distributions might also be used, such as the Laplace distribution.}

\subsubsection{Practitioner meets statistician}\label{Sec:practitioner.meets.statistician}

If we decide to use $\mfP_1(z_t)=G(z_t)$, \eqref{ML.update.full} will be similar to \eqref{Elo.practitioner.compact}. Both will be identical when $\phi(z) \propto 1$ which, from \eqref{xi.definition}, is satisfied if
\begin{align}
\label{differential.equation}
    \dot\mfP_1(z)\propto[1-\mfP_1(z)]\mfP_1(z).
\end{align}
This differential equation is uniquely solved\footnote{To be more precise, $\mfP_1(z)=\frac{1}{1+\e^{-z/s+b}}$ solves \eqref{differential.equation} for any $b\in\Real$.} by the logistic function
\begin{align}\label{P1=mcL}
    \mfP_1(z) &= \mcL(z/s)\\
    \label{logistic.definition}
    \mcL(z)&=\frac{1}{1+\e^{-z}}.
\end{align}

Thus, the only way to obtain the equivalence between the practitioner's and the statistician's rankings is by using $G(z)=\mcL(z)$. Indeed, this choice is often made in practice, and then, the practitioner's approach is endowed with a clear statistical interpretation of the ranking results: 
if we set $G(z)=\mcL(z)$, the Elo algorithm solves the \gls{ml} estimation problem using the \gls{sg} algorithm.

What happens when we set $G(z)$ arbitrarily? This is a valid question because, in the original definition of the Elo ranking in \cite[Sec.~1.4]{Elo78_Book}, the expected score is defined as
\begin{align}
\label{G(z)=Phi(z)}
    G(z)=\Phi(z),
\end{align} 
where $\Phi(z)=\frac{1}{\sqrt{2\pi}}\int_{-\infty}^z\exp\left(-0.5x^2\right)\dd x$ is the Gaussian \gls{cdf}

In this case, by setting $\mfP_1(z)=G(z)=\Phi(z)$, we obtain $\phi(z)\centernot{\propto}1$; thus, the Elo algorithm implements only \emph{approximately} the \gls{ml}+\gls{sg} estimation. However, since the \gls{sg} is also an approximate solution, in practice, this distinction is not critical.

In fact, our purpose is not to exaggerate the importance of the statistician's perspective but rather to harness it to obtain a clear interpretation of the results that are obtained by practitioners. Some differences between the approaches may be subtle: the practitioner may ignore the existence of probabilistic model which leads to difficulties in interpretation -- especially for matches with multi-level outcomes. Other differences are more important: the estimation errors are ignored by practitioners while the statistician explicitly assumes that $\btheta_t$, as the estimates of the skills $\btheta^*$, are subject to estimation errors which occur due to use of the \gls{sg} and/or due to finite number of matches.

And finally, we note that the distinction between both approach may be sometimes blurred: the description of the practitioner's approach from Sec.~\ref{Sec:Practitioner.perspective} refers to the Elo algorithm, but the original work of 
Arpad Elo \cite{Elo78_Book}, did not entirely ignore the statistical modelling. In fact, \cite{Elo78_Book} used the Thurstone model of the pairwise comparisons \cite{Thurston27} (the work, which is, notably, almost 100 years old!) to derive the model \eqref{G(z)=Phi(z)}.\footnote{On the other hand, despite the probabilistic model, the estimation method, that is the Elo algorithm itself, as explained in \cite{Elo78_Book} does not appeal to any statistical method (such as \gls{ml}) and belongs rather to the heuristic domain, \ie the practitioner's one. This is particularly obvious when we consider matches with non-binary outputs for which the original Elo algorithm was devised without specifying the probabilistic model at all. More on that in Sec.~\ref{Sec:Multilevel.outcomes}.}

However, we believe that it was the practitioner's approach laid out by Arpad Elo in \cite{Elo78_Book}, thanks to the intuitive explanation and a simple operation of the algorithm, which made this ranking algorithm so popular. In fact, the statistician's approach was explicitly shown much later, \eg in \cite{Kiraly17} and \cite{Szczecinski20}.

\subsubsection{Re-scaling and HFA}\label{Sec:Scaling.HFA}

The estimated skills in the vector $\btheta_t$ may be quite small,\footnote{The reason is that the match results are not extreme, for example, $\mfP_1(z)=\mcL(z)\in(0.2,0.8)$ which corresponds to $z\in(-1.4,1.4)$ and $\theta_{t,m}\in(-0.7,0.7)$.} and it is customary to multiply them by the scale $s>1$, \ie we make the replacements $\btheta_t \leftarrow s \btheta_t$ and $z_t\leftarrow s z_t$, which integrated into the recursive estimation \eqref{Elo.practitioner.compact} produce
\begin{align}
\label{Elo.practitioner.compact.scale}
    \btheta_{t+1} 
    &= \btheta_t + \mu s \bx_t\big[y_t - G(z_t/s) \big]\\
\label{Elo.practitioner.compact.scale.2}
    &= \btheta_t + K \bx_t\big[y_t - G(z_t/s) \big]
\end{align}
where, in the conventional presentation of the Elo algorithm shown in \eqref{Elo.practitioner.compact.scale.2}, the scale is absorbed into the step $K=\mu s$. However, leaving the scale in the notation \eqref{Elo.practitioner.compact.scale}, clarifies that by changing the scale from $s$ to $\tilde{s}=\beta s$, in order to maintain the same algorithm's behavior we should also change the step from $K$ to $\tilde{K}=\beta K$.\footnote{Then, denoting by $\tilde\btheta_t$ the skills obtained using the scale $\tilde{s}$ and the step $\tilde{K}$, we have $\tilde\btheta_t=\beta\btheta_t$, where $\btheta_t$ are the skills obtained in \eqref{Elo.practitioner.compact.scale.2}.}

Although the scaling is arbitrary, it is useful when comparing different ranking algorithms: here we treat $G(z)=\mcL(z)$ as a ``canonical'' form of the expected score in the Elo ranking, but other functions $G(z)$ may be similar to $\mcL(z)$ after appropriate scaling, \ie $G(z/\tilde{s})\approx \mcL(z/s)$.

For example, using, as the expected score, a generalized logistic function  
\begin{align}
    G(z)=\mcL(z;a)=\frac{1}{1+a^{-z}}   
\end{align}
with an arbitrary exponential basis, $a>0$, we can write
\begin{align}
\label{from.e.to.decimal}
    \mcL(z/\tilde{s};a) &= \mcL(z/s),\\
\label{from.e.to.decimal.tilde.s}
    \tilde{s} &=s \beta_{\e\rightarrow a},
\end{align}
where we use a mnemonic notation
\begin{align}
    \beta_{\e\rightarrow a} = \log a
\end{align}
to indicate the scale adjustment required to replace the canonical logistic function $\mcL(\cd)$ by a generalized logistic function $\mcL(\cd;a)$. In this particular case, after the scaling \eqref{from.e.to.decimal}, we obtain not only similar, but identical ranking algorithm regardless of the basis $a$. Similarly, we can go from $\mcL(\cd;a)$ to $\mcL(\cd)$ using $\beta_{a\rightarrow\e}=1/\beta_{\e\rightarrow a}$.

If we decide to use $G(z)=\Phi(z)$, the expected scores (and the ranking algorithms) can be made only approximately equivalent
\begin{align}\label{from.Phi.to.L}
    \Phi(z/\tilde{s}) \approx \mcL(z/s).
\end{align}
For example, 
the ``equivalence'' may be enforced by making the derivatives equal for $z=0$, \ie
\begin{align}
\label{equal.2nd.derivatives.Binary}
    \frac{1}{\tilde{s}}\dot\Phi(z/\tilde{s})|_{z=0}
    &=
    \frac{1}{s}\dot\mcL(z/s)|_{z=0},
\end{align}
where $\dot\Phi(z)=\frac{\dd}{\dd z}\Phi(z)=\mcN(z)=\frac{1}{\sqrt{2\pi}}\exp(-0.5z^2)$, and $\dot\mcL(z)=\mcL(z)\mcL(-z)$. This yields
\begin{align}
    \tilde{s} &= s \beta_{\mcL\rightarrow\Phi},\\
\label{beta.L.2.Phi}
    \beta_{\mcL\rightarrow\Phi} &= \frac{4}{\sqrt{2\pi}}\approx 1.6.
\end{align}
Although somewhat arbitrary,\footnote{
An alternative, also used in the literature, \eg \cite[Appendix~A]{Glickman99}, \cite[Sec.~5.1.1]{Szczecinski23a} equates the variances of the logistic and Gaussian distributions
\begin{align}
    \frac{s^2\pi^2}{3} &= \tilde{s}^2,\\
    \label{beta.L.2.Phi.moments}
    \beta_{\mcL\rightarrow\Phi} &=\frac{\tilde{s}}{s}=\frac{\pi}{\sqrt{3}}\approx 1.8.
\end{align}
}
this approximation principle has value of being simple.

In fact, most of the current versions of the Elo algorithm use the logistic functions $\mcL(z)$ or $\mcL(z;10)$, however, the transformation between $\Phi(z)$ and $\mcL(z)$ is useful in approximate calculations we will see in Sec.~\ref{Sec:Performance evaluation}.

\textbf{\Gls{hfa}}\\
In practice, it is often observed that playing at home (that is at team's stadium or player's country) provides an ``advantage'' to the player/team. Depending on the sport, this is known as home-field/courts/ice/turf/ground advantage.

The \gls{hfa} effect is modeled by increasing the difference $z_t$
\begin{align}
\label{Elo.practitioner.compact.scale.HFA}
    \btheta_{t+1} &= \btheta_t + K \bx_t\big[y_t -G(z_t/s+\eta h_t) \big]\\
    &=\btheta_t + K \bx_t\big[y_t -G\big((\theta_{t,i_t} - \theta_{t,j_t}+ \eta s h_t )/s\big) \big]
\end{align}
where $\eta$ is the scale-independent \gls{hfa} factor and $h_t=1$ indicates that the match is played on the player's home venue (country/arena/stadium) and $h_t=0$ indicates the neutral venue (\eg during competitions hosted by a country different from the home- and away players). To simplify the notation we assume henceforth $h_t=1$.

Formulation in \eqref{Elo.practitioner.compact.scale.HFA} corresponds to boosting the skills of the home player by $s \eta$, which, alternatively may be seen as a value by which the skills of the away player must be increased to neutralize the \gls{hfa} effect. The meaning of the \gls{hfa} can be naturally extended to any identifiable element that favors the home player. For example, in chess, the player starting with white pieces has, on average, better chances of winning.

Similarly, in the statistician's approach, we use
\begin{align}
\label{P1.HFA}
    \PR{Y_t=1|z}&=\mfP_1(z_t/s+ \eta),
\end{align}
and can also directly modify the algorithm \eqref{ML.update}
\begin{align}
    \btheta_{t+1} 
\label{ML.update.scale.HFA}
    &= \btheta_{t} + \mu s \bx_t \dot\ell_{y_t}(z_t/s +\eta).
\end{align}

If we want to change the expected score, \eg from $G(z/s)$ to the canonical logistic function $\mcL(z/\tilde{s})$ this is done as follows:
\begin{align}\label{from.G.to.L}
    G(z_t/s+\eta)
    &=G\big((z_t+\eta s)/s\big)\\
    &\approx
    \mcL\big((z_t+\eta s)/\tilde{s}\big)
    =\mcL\big((z_t+\tilde\eta \tilde{s})/\tilde{s}\big)\\
    &=\mcL\big(z_t/\tilde{s}+ \tilde{\eta} \big),\\
    \tilde{s}&=s \beta,\\
\label{tilde.eta.from.eta}
    \tilde{\eta}&=\eta/\beta,
\end{align}
where 
$\beta$ is the scale adjustment factor, \eg $\beta=1/\beta_{\e\rightarrow a}$ or $\beta=1/\beta_{\mcL\rightarrow\Phi}$ (depending on whether we go to $\mcL(z/\tilde{s})$ from $G(z/s)=\mcL(z/s;a)$ or $G(z/s)=\Phi(z/s)$) and $\tilde{\eta}$ is the new \gls{hfa} term to be used in the canonical logistic score, and we preserve the product $\eta s = \tilde\eta \tilde{s}$. 

For example, $\eta s=100$ and $s=400$ in \cite{eloratings.net} (\ie $\eta=0.25$) which uses $G(z)=\mcL(z;10)$ so, by moving the algorithm to the canonical form $\mcL(z)$, we should use $\beta=1/\beta_{\e\rightarrow 10}=0.43$ and $\tilde{s}=s\beta\approx 174$ with $\tilde{\eta}=0.25/\beta \approx 0.58$ or $\tilde\eta\tilde{s}=100$.  

Thus, if we want to rescale the skills in the same algorithm (the same expected score), the term $\eta$ should be kept separated from the scale, \ie the notation $G(z/s+\eta)$ is useful, but if we change the function defining the expected score, the product $\eta s$ is preserved across the functions.
\subsubsection{Convergence}\label{Sec:Convergence}

The detailed description of the dynamics defined through \eqref{Elo.practitioner.compact.scale} is rather complex, see \cite{Jabin20}, \cite{Cortez26} but it is analyzed in \cite{Zanco24} in terms of the average variance of the skills estimate after convergence 
\begin{align}
\label{average.variance.definition}
    \ov{v}&=\lim_{t\rightarrow\infty} \ov{v}_t,\\
    \ov{v}_t&=\frac{1}{M}\sum_{m=1}^{M}\Ex[(\theta_{t,m}-\Ex[\theta_{t,m}])^2],
\end{align}
where the following approximation is proposed \cite[Eq.(50)]{Zanco24} 
\begin{align}\label{average.variance.from.K}
    \ov{v}&\approx s^2\mu/2=sK/2.
\end{align}

If the outcomes $y_t$ are generated using the model \eqref{P1.HFA} with the scale $s^*$, we may assume that $\Ex[\theta_{\infty,m}]=\theta_m^* s/s^* $, that is, the ground truth must be rescaled; this will be taken into account and, unless stated otherwise, we use $\theta_m^*\equiv \theta_m^* s/s^*$. The variance $\ov{v}$ has a theoretical meaning because it can be measured only with multiple realizations of $\theta_{t,m}$ which exist only in simulations; in practice, it can be calculated by temporal averaging as
\begin{align}
    \ov{v}_t &\approx \frac{1}{W}\sum_{\tau=-W+1}^t |\theta_{\tau,m}-\ov{\theta}_{t,m}|^2,\\
    \ov{\theta}_{t,m} &\approx \frac{1}{W}\sum_{\tau=-W+1}^t \theta_{\tau,m},
\end{align}
where $W$ is the averaging window length.

Moreover, \cite[Sec.~3.3]{Zanco24} tells us that the skills converge in expectation as 
\begin{align}\label{convergence.in.expectation}
    \Ex[\theta_{t',m}]\approx\e^{-t'/\tau}\big(\theta_{0,m}-\Ex[\theta_{\infty,m}]\big) + \Ex[\theta_{\infty,m}],
\end{align}
where $t'=0,1,\ld$ are indices of the matches in which the player $m$ is participating, and the convergence time constant is given by
\begin{align}\label{tau.from.K}
    \tau=\frac{4}{\mu}=\frac{4s}{K}.
\end{align}
In \cite{Zanco24}, a round-robin tournament with one game per time index is analyzed, so the players, on average, wait $\ov{T}=M/2$ games before playing again. In such a case, the time constant for all the games must be multiplied by $\ov{T}$.

In a practice, it is common to use variable adaptation step $K_t$ (defined, \eg in \gls{fifa} ranking, according to the importance of the match \cite{fifa_rating}, or, in \gls{fivb} ranking, according to the number of matches played). In such a case the variance and the time constant are obtained using the average adaptation step
\begin{align}\label{average.tau}
    \ov{\tau}_m &= \frac{4s}{\ov{K}_m},\\
    \ov{v} &= \frac{1}{M}\sum_{m=1}^M \frac{s\ov{K}_m}{2},\\
    \ov{K}_m &= \frac{1}{N_m}\sum_{\substack{t\\i_t=m \vee j_t=m}} K_t,
\end{align}
where $N_m$ is the number of matches played by player $m$. 

By a rule of thumb, after $2\ov\tau-3\ov\tau$ matches, the convergence in expectation may be declared \cite[Sec.~4.2]{Zanco24}. Thus, for a given scale $s$, increasing $K$ (or $K_t$), we increase the convergence speed (the convergence time is inversely proportional to $K$) at the expense of the larger estimation errors after convergence (the variance grows linearly with $K$). We illustrate this in Example~\ref{Example:Illustration.convergence}.

\begin{example}[Illustration of convergence]
\label{Example:Illustration.convergence}
We show in Fig.~\ref{fig:Elo.binary_games} an example obtained by simulations, where $M=30$ players with skills $\theta^*_m$ (generated using Gaussian distribution with variance $v_\theta=0.5$) compete in a round-robin fashion with random matchups: at time $t$ each player is matched with a randomly chosen opponent (this is a random $
\bx_t$) and the outcome of the match $y_t$ is generated with $\mfP_1(z)=\mcL(\bx\T_t\btheta^* + \eta^*)$, where $\eta^*=0.35$; thus, the scale is $s^*=1$.

The Elo algorithm uses the canonical expected score $\mcL(z/s+\eta)$  with $\eta=\eta^*$ and the scale $s=174$.\footnote{This is inspired by the \gls{fide} ranking based on  $\mcL(z;10)$ and $s=400$ which, after applying \eqref{from.e.to.decimal} yields $\tilde{s}\approx174$} 
We consider two cases of adaptation step a) fixed on $K=60$, and b) random $K_t$ selected with uniform probability from the set $\set{10, 20, 30}$, thus we have $\ov{K}_m=20$.

For two selected players, the trajectories of their skills $\theta_{t,m}$ are shown with thin lines (blue when $K=20$ and green when $K=60$) for $J=200$ different realization of $\bx_t$ and $y_t$. The thick red curve indicates the analytical mean $\Ex[\theta_{t,m}]$, the blue marker -- the empirical mean, and the dashed red lines -- the mean plus/minus standard deviation $\Ex[\theta_{t,m}]\pm\sqrt{\ov{v}}$ (which is a rudimentary credible interval for $\theta_{t,m}$). The analytical formulas give a reasonable prediction for a constant step $K$ and random $K_t$.

Using random $K_t$, the average time constant \eqref{tau.from.K} is equal to $\ov\tau=4\tilde{s}/\ov{K}\approx 4\cd174/20 \approx 35$, which, by a rule of thumb, means that after $2\tau-3\tau\approx 70.-104.$ matches, the convergence in expectation may be declared. For $K=60$, $2\tau-3\tau\approx 35-50$. The differences in time constants can be appreciated in the behavior of the curves shown in Fig.~\ref{fig:Elo.binary_games}.

The variances \eqref{average.variance.from.K} $\ov{v}=\ov{K}\tilde{s}/2\approx 1740$  (for $\ov{K}=20$), and $\ov{v}=5220$ (for $K=60$) were reasonably close to the corresponding values $\approx1762$ and $\approx5807$ we obtained empirically.

\begin{figure}
    \centering
    \includegraphics[width=0.9\linewidth]{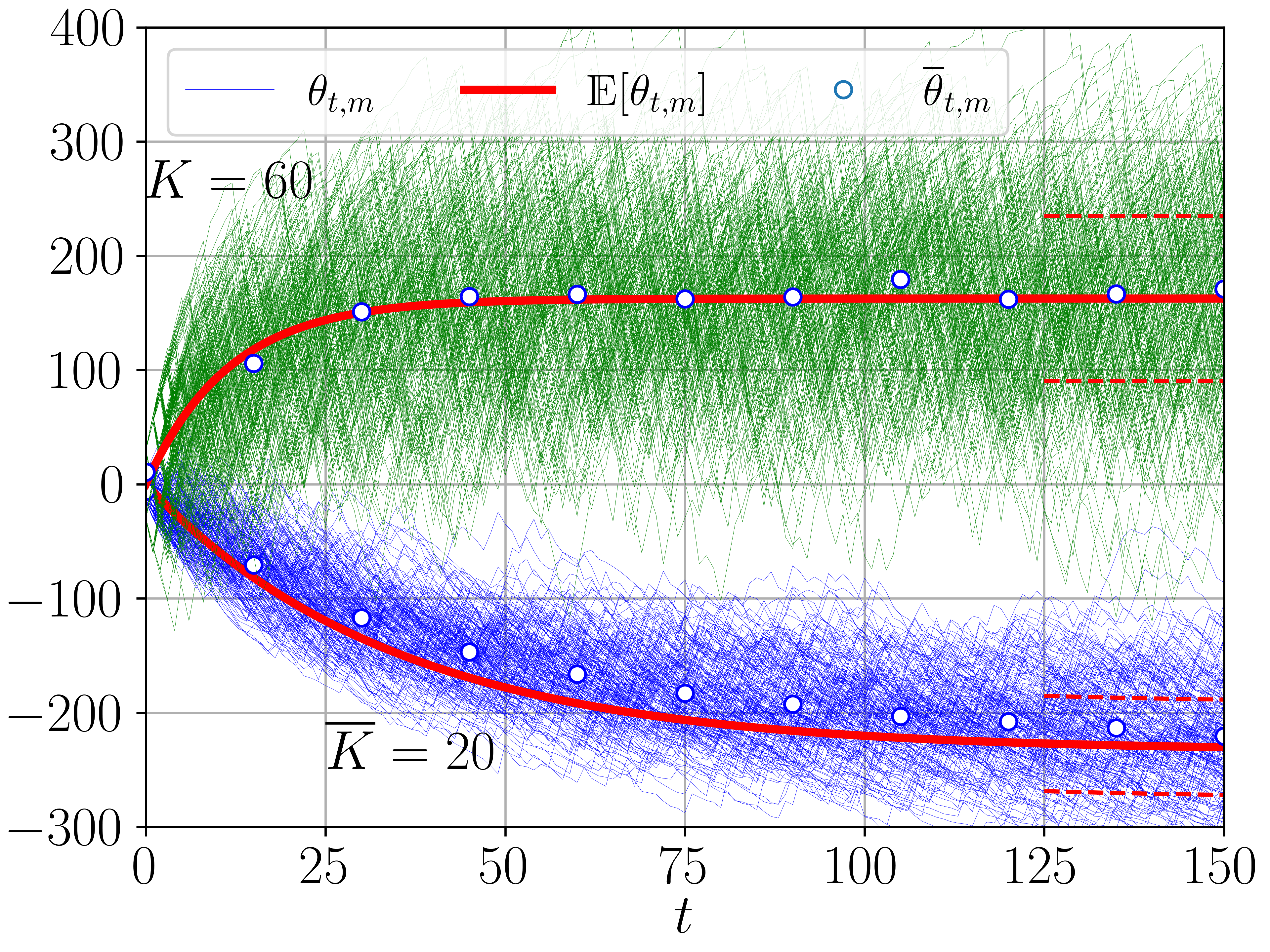}\\
    \caption{Trajectories of the estimated skills obtained using the Elo algorithm with random step $K_t\in\set{10,20,30}$ (lower, blue) or with fixed $K=60$ (upper, green). Done for a given $\btheta^*$, each curve corresponds to a different realization of $y_t$ and $\bx_t$. Markers indicate the empirical average (across realizations) and the dashed red lines indicate the 68\% credible interval at convergence ($\Ex[\theta_{\infty,m}]\pm \sqrt{\ov{v}}$).}
    \label{fig:Elo.binary_games}
\end{figure}
\end{example}

\subsubsection{Interpretation of the skills}\label{Sec:Performance evaluation}

\textbf{Who is better?}\\
If we need the ranking algorithm to find which team/player is stronger/better, it should be done comparing the estimated skills.

For the practitioner, there is no problem here: $\theta_{t,m}>\theta_{t,n}$ means that the team $m$ is better than the team $n$. The statistician, on the other hand, is aware that the skills $\theta_{t,m}$ are corrupted by the estimation errors $\xi_{t,m}$, \ie $\theta_{t,m}=\theta^*_{t,m} + \xi_{t,m}$, 
and thus
\begin{align}\label{zt.z*t}
    z_t=z^*_t + \zeta_{t},
\end{align}
where $z^*_t=\theta^*_{t,i_t}-\theta^*_{t,j_t}$ is the difference between the true skills, and $\zeta_t=\xi_{t,i_t}-\xi_{t,j_t}$ is the difference between the estimation errors.  

As a consequence of \eqref{zt.z*t}, the question ``is $m$ better than $n$?'' (\ie is $z^*_t>0$?) can only have a probabilistic answer by conditioning on the observed difference $z_t>0$
\begin{align}
    \PR{z^*_t>0|z_t>0}
    &=
    \PR{z_t>0|z^*_t>0}\\
    &=\PR{\zeta_t>-z_t}.
\end{align}

If we assume that $\xi_t$ are \gls{iid} Gaussian variables\footnote{This is a simplifying assumption because skills and the corresponding errors are obtained from the same outcomes; thus, they are not independent.} with zero-mean and variance $\ov{v}$, then
\begin{align}
    \PR{z^*_t>0|z_t>0}&=\Phi(z_t/\sqrt{2\ov{v}})
    \\
    &=
    \begin{cases}        
        0.69 & z_t = 0.5 \sqrt{Ks}\\
        0.84 & z_t = \sqrt{Ks}\\
        \ld
    \end{cases}
\end{align}

Thus, by a rule of thumb, we may require, before confidently declaring one player superior to the other, the difference between the skills should be at least larger than $\sqrt{Ks}$. In Example~\ref{Example:Illustration.convergence}, we have $\sqrt{\ov{K}s}=90$ (for $\ov{K}=20$), and $\sqrt{Ks}=104$ (for $K=60$). For comparison, the \gls{fide} ranking (where the smallest step is $K=16$ and the canonical scale is the same as in Example~\ref{Example:Illustration.convergence}) published in March 2026, shows that the differences between the skills at consecutive positions (for the first 20 places) are smaller than $20$[skills points]. Then, declaring confidently who is the best is not really possible!

\textbf{Prediction/performance evaluation}\\
Prediction of the matches results from the previous observations is reduced to the prediction from the skills, that is we want to find
$\PR{Y_\tau=y|\btheta_t}$ for any $\tau\ge t$. 

This may be useful to analyze the structure of the tournament, \eg \cite{Csato23}\cite{Lapre25}, to evaluate the betting odds \cite{Egidi18}, to predict the outcomes of a competition \cite{Brandes25}, or to compare the quality of the ranking models/algorithms \cite[Sec.~2]{Gelman14}.

In these cases, the relevant metric is the log-likelihood of the future matches $\ell_{y_\tau}(z_t)=\log\PR{Y_\tau=y_\tau|\btheta_t}$ which may be averaged across testing set of matches with indices in $\mcT^\tnr{test}$ yielding the mean log-score
\begin{align}\label{neg.log.scores.definition}
    \mf{LS} = -\frac{1}{|\mcT^\tnr{test}|}\sum_{\tau\in\mcT^\tnr{test}} \ell_{y_\tau}(\bx\T_\tau \btheta_t/s+\eta),
\end{align}
where negation is optional but yields a positive log-scores with lower value meaning a ``better'' prediction. The main formal requirement is that $\btheta_t$ not be calculated using $y_\tau$.  In the Elo ranking $y_t$ affects $\btheta_{t+1}$, so we can use the entire data set for testing, \ie $\mcT^\tnr{test}=\mcT$. 

The task at hand seems simple for the practitioner: it is enough to use the model which appeared in Sec.~\ref{Sec:Practitioner.perspective} and calculate
\begin{align}\label{Pr1.practitioner}
    \ell_{y_t}(z_t) &= \log\mcL(z_t/s+\eta).
\end{align}

For the statistician, on the other hand, the model \eqref{P1=mcL} connects the true skills $\theta^*_m$ to the outcomes $y_t$, so, the relationship to the noisy skills $\theta_{t,m}$ (and thus also to $z_t$) can be found through marginalization over $z^*_t$
\begin{align}\label{ov.Pr.Y=1}
    \ov\mcL(z^*_t/s+\eta)&=\int \mcL(z^*_t/s+\eta)\pdf(z^*_t|z_t)\dd z^*_t
\end{align}
where
\begin{align}\label{pdf.z^*.z}
    \pdf(z^*_t|z_t)\propto\pdf(z_t|z^*_t)\pdf(z^*_t)/\pdf(z_t),
\end{align}
that is, we model the true $z^*_t=\theta^*_{t,i_t}-\theta^*_{t,j_t}$ as random variables which makes sense because $i_t$ and $j_t$ are randomly selected and it is common to assume that the skills $\btheta^*_t$ are realizations of the random variables drawn from a Gaussian distribution with variance $v_\theta$. Then, $z^*_t$ may be thought about as a realization of a Gaussian variable with variance $2 v_\theta$, \ie $f(z^*_t)=\mcN(z^*_t;0, 2v_\theta)$.

To calculate \eqref{ov.Pr.Y=1} in a closed form, we assume again that $\xi_{t,m}$ are zero-mean Gaussian variables with variance $\ov{v}$, \ie $\pdf(z|z^*)=\mcN(z;z^*,2\ov{v})$ and then, as shown in Appendix~\ref{Proof.empirical.expected.score}, we have
\begin{align}\label{formula.empirical.expected.score}
    \ov\mcL(z^*_t/s+\eta)
    &\approx
    \mcL\left(z_t/\hat{s}+\hat\eta\right)
    =\mcL\left((z_t+\eta s a)/\hat{s}\right)
\end{align}
where
\begin{align}
\label{s.hat.from.s}
    \hat{s} &= s\beta_\tnr{err},\\
\label{hat.eta.transformed}
    \hat\eta &= \frac{\eta a}{\beta_\tnr{err}},
\end{align}
are the effective scale and \gls{hfa},
\begin{align}
\label{tilde.beta.binary}
    \beta_\tnr{err}&=a\sqrt{1+\frac{2\ov{v}}{a( s\beta_{\mcL\rightarrow\Phi} )^2}},
\end{align}
is the scale adjustment factor due to the estimation errors, 
$a=(1+\ov{v}/v_\theta)$ is defined in \eqref{a.definition}, and $\beta_{\mcL\rightarrow\Phi}$ is the scale adjustment \eqref{beta.L.2.Phi} used for model equivalence in Appendix~\ref{Proof.empirical.expected.score}.

The average log-score \eqref{neg.log.scores.definition} is then calculated as 
\begin{align}\label{neg.log.scores.definition.corrected}
    \mf{LS} = -\frac{1}{|\mcT|}\sum_{t\in\mcT} \ell_{y_t}(z_t/\hat{s}+\hat\eta).
\end{align}

Only if we neglect the estimation errors, \ie when $\ov{v}\approx 0$, we have $\beta_\tnr{err} \approx 1$. This is what happens in \eqref{Pr1.practitioner} because, in the practitioner's perspective, the estimation errors are not present.

\textbf{Pseudo model-identification}\\
Since the analytical formulas rely on assumptions and unknown parameters: $v_\theta$, Gaussian errors $\zeta_t$, etc, we may prefer to find the parameters $\hat{s}$ and $\hat\eta$ from the data directly through the following optimization:
\begin{align}\label{hat.s.from.data}
    \hat{\gamma}, \hat\eta &=\argmax_{s,\eta} \sum_t \ell_{y_t}(\gamma z_t/s+\eta),\\
    \hat{\beta}&= 1/\hat\gamma;
\end{align}
for given $z_t$, the function under optimization is concave in $\eta$ and in $\gamma$ (but not concave in $\beta=1/\gamma$).

While this is a formulation typical in the \gls{ml} model identification problems, we are \emph{not} identifying the model per se but rather the joint effect of (i) the estimation noise in $z_t=\bx\T_t\btheta_t$,  (ii) the bias introduced by the limited variance $v_\theta$, and of (iii) the mismatch between the data and the models used in the estimation/ranking and the performance evaluation (in this synthetic example, there is no mismatch: all models are logistic).

After convergence, we always have $\beta_\tnr{err}>1$, thus the effective scale $\hat{s}$ increases with the noise in the skills' estimates, $\ov{v}$, but also with the ratio $\ov{v}/v_\theta$. Recognizing the latter, differs from the literature that often assumes only $\ov{v}$ affects the prediction (\eg  \cite[App.~E]{Ingram21}), which is true if $v_\theta$ is sufficiently large, \ie if the dispersion of the skills $\btheta_t^*$ is much larger than the variance of the estimation errors; then $\pdf(z^*_t|z_t)=\pdf(z_t|z^*_t)$.

\begin{example}[Fit to the estimated data]\label{Ex:Pseudo.model.ident.binary}
Let us reuse Example~\ref{Example:Illustration.convergence} to find the scales through a fit to the estimated data where $T=2000$ last matches after the convergence were used.

The results are shown in Table~\ref{tab:logscores}, where we see that (a) with growing estimation errors the scale $\hat{s}$ increases and the effective \gls{hfa} $\hat\eta$ decreases, (b) the theoretical formulation in \eqref{s.hat.from.s} explains quite well the results obtained through the actual fit to the data \eqref{hat.s.from.data}, and that (c) the scale fitting does not provide any meaningful gain for $K=20$, but an improvement can be seen for $K=60$, \ie in the context of larger estimation errors.


\begin{table}[]
\caption{Log-score \eqref{neg.log.scores.definition.corrected} computed using the practitioner approach (assuming no estimation error),
the theoretical approach \eqref{s.hat.from.s} and \eqref{hat.eta.transformed}, and the
data-fitting approach based on \eqref{hat.s.from.data}.
Here $s=174$, $\eta^*=0.35$, $M=30$, $T=1000$. Values in parentheses denote (mean, std) obtained from $J=200$ realizations.}
\centering
\resizebox{\columnwidth}{!}
{
\begin{tabular}{lccc ccc}
\toprule
& \multicolumn{3}{c}{$K=20$} & \multicolumn{3}{c}{$K=60$} \\
\cmidrule(lr){2-4}\cmidrule(lr){5-7}
Method & $\hat{\beta}$ & $\hat{\eta}$ & $\mathsf{LS}$ & $\hat{\beta}$ & $\hat{\eta}$ & $\mathsf{LS}$ \\
\midrule
Assume no error  & $1$                & $\eta^*$           & $(0.59, 0.01)$ & $1$                & $\eta^*$           & $(0.62, 0.01)$ \\
Theoretical      & $1.10$             & $0.34$             & $(0.59, 0.01)$ & $1.40$             & $0.33$             & $(0.60, 0.01)$ \\
Data-fit         & $(1.13, 0.04)$     & $(0.34, 0.04)$     & $(0.59, 0.01)$ & $(1.45, 0.05)$     & $(0.33, 0.04)$     & $(0.60, 0.01)$ \\
\bottomrule
\end{tabular}
}

\label{tab:logscores}
\end{table}

\end{example}

The formula \eqref{tilde.beta.binary} explains quite well why the actual fit to the data via \eqref{hat.s.from.data} produce $\hat{s}>s$ and $\hat\eta<\eta^*$. On the other hand we know that the discrepancies may be produced by the estimation error and/or the distribution of the skills as defined via $v_\theta$ and any other mismatch between the model and the data. What finally counts, is how to estimate the performance and, using  $\hat{s}$ and $\hat\eta$ provides a simple answer. In that sense, \eqref{hat.s.from.data} seems to be a more practical approach  which does not need any assumptions required to derive \eqref{average.variance.from.K}.

Another reason to use \eqref{hat.s.from.data} is that \eqref{tilde.beta.binary} works after convergence.  On the other hand, before convergence is attained, the skills may be ``compressed''  when initialization $\theta_{0,m}$ is smaller than $\theta_{0,m}^*$. In this case the skills differences may be too small, \ie $|z_t|<|z^*_t|$. Then, the theoretical formulas fail, but the empirical fit will unveil the lack of convergence by decreasing the scale $\hat{s}$, that is, producing $\hat\beta<1$.

As shown in Table~\ref{tab:logscores}, in some cases, the scale fitting may appear superfluous. However, it is not a costly operation and is helpful when dealing with large estimation errors. More importantly, it introduces a key idea that the model used for ranking need not be the same as the one used for prediction. This conceptual model ``decoupling'' will be necessary when the model underlying the Elo algorithm is not explicitly specified; more on that in Sec.~\ref{Sec:model.decoupling}.

Note that instead of pseudo-fitting the model, we may also find the expected score $\mcL(z/\hat{s}+\hat\eta)$ from the data: this was done in  \cite{Sonas11} via histogram of the score $\rho_{y_t}$, which found the scale $\hat{s}\approx 1.25 s$. However, \cite{Sonas11} did not explain that phenomenon, most likely because, adopting the practitioner's perspective, the estimation errors were simply not taken into account.

\textbf{Dispelling potential confusion}\\
The fact that we can have the ``nominal'' expected score $\mcL(z/s+\eta)$ and the empirical expected score $\mcL(z/\hat{s}+\hat\eta)$ may lead to a confusion: which one ``true"? In fact, both are true/correct! The empirical one allows us to calculate the expected score from the estimated skills difference $z=z_t$ (corrupted by the estimation errors $\zeta_t$) while the nominal one, $\mcL(z/s+\eta)$ relates the expected score to the noise-free skills difference $z=z^*_t$.\footnote{While it might be tempting to use the ``true'' expected score $\mcL(z/\hat{s}+\hat\eta)$ in the Elo algorithm, it would be pointless and would merely rescale the skills. In fact, to take into account the knowledge of the fact that the estimated skills are corrupted by errors, the estimation problem has to be modified resulting in  the simplified form of a Kalman filter. However, showing these details would take us too far from the subject of this work so we refer the interested readers to \cite{Ingram21} and \cite{Szczecinski23a}.}

\section{Beyond win/loss matches: Elo ranking for multilevel outcomes}\label{Sec:Multilevel.outcomes}

\subsection{Practitioner's perspective: Simplicity!}\label{Sec:multilevel.practitioner}

The principles underlying the practitioner's version of the Elo algorithm are so simple that the algorithm is also used when matches have multilevel outcomes, \ie when $y_t\in\mcY=\set{0,\ld, L-1}$ and $L>2$. This is done ``naturally'' so the algorithm works as before, \ie \eqref{Elo.practitioner.compact.scale.HFA}
\begin{align}
\label{Elo.rating.multilevel}
    \btheta_{t+1}&=
    \btheta_{t} + K \bx_t [\rho_{y_t} - \mcL(z_t/s+\eta h_t)],
\end{align}
where, to focus the analysis, we use the logistic function \eqref{logistic.definition} as the expected score, \ie $G(z)=\mcL(z)$. 

To apply the algorithm, we only need to assign numerical scores to the match's results, \ie $ \rho_y\in[0,1], y=0,\ld,L-1$ where we use $\rho_0=0$ to denote the least important result and $\rho_{L-1}=1$ denotes the most important one.\footnote{For example, in volleyball, where $L=6$, we would use $y_0$ to denote the home team losing ``0-3'', and $y_5$ to denote the home team winning ``3-0''.} The scores $\rho_y$ will be monotonic in $y$, as the practitioner will assign a higher score to a more important outcome $y$; we store them in a vector $\brho=[\rho_0,\ld, \rho_{L-1}]$.

In fact, this practitioner's perspective has been the basis for the Elo algorithm devised to rate chess players \cite{Elo78_Book}, where matches can end in a draw (pat), for which the score was defined as $\rho_1=0.5$ but where no probabilistic model was specified for three possible match outcomes.

Regarding the model, we recall that, at the end of Sec.~\ref{Sec:Practitioner.perspective},  for $L=2$, the ``expected score'' was defined using the mathematical expectation interpretation, \ie
\begin{align}
\label{G(z).definition}
    G(z) = \sum_{y=0}^{L-1} \mfP_y(z) \rho_y.
\end{align}

For $L=2$, \eqref{G(z).definition} implies a unique probabilistic model, $\mfP_1(z)=G(z)$ and $\mfP_0(z)=1-G(z)$. However, for $L>2$, there is no such relationship and this leads to the following question: What probabilistic model, if any, underlies the algorithm \eqref{Elo.rating.multilevel}?

To answer it, we will continue  treating the Elo ranking \eqref{Elo.rating.multilevel} as the \gls{ml}+\gls{sg} optimization \eqref{ML.update.scale.HFA} based on the derivative of unknown log-likelihood $\tilde{\ell}_y(z)$, that is,
\begin{align}
    \label{Elo.rating.multilevel.derivative}
    \btheta_{t+1}&=
    \btheta_{t} + \mu s \bx_t \dot{\tilde{\ell}}_y(z/s+\eta),
\end{align}
where
the derivative is given in \eqref{Elo.rating.multilevel}
\begin{align}
    \dot{\tilde{\ell}}_y(z/s+\eta) =  b \big[\rho_y-\mcL(z/s+\eta)\big],
\end{align}
where $b>0$ is a multiplying constant that is part of step $K=\mu b s$, and we assume $h_t=1$.

We find $\tilde{\ell}_y(z)$ by integration removing for convenience the scale, $s$, and the \gls{hfa}, $\eta$,
\begin{align}
\label{ell.from.integration}
    \tilde\ell_y(z) 
    &= b\int_{-\infty}^{z} \rho_y-\mcL(u) \dd u\\
\label{ell.from.integration.result}
    &= \rho_y b z - b\log(1+\e^{z}) +c_y,
\end{align}
where $c_y>-\infty$ is an arbitrary constant, and the model is then found as\footnote{We excluded the case of $c_y=-\infty$ which yields $\mfP_y(z)\equiv 0$ irrespectively of $\rho_y$ and $z$.} 
\begin{align}
\label{Py.z.identified}
     \tilde\mfP_y(z)&=\e^{\tilde\ell_y(z)}
     = \frac{\e^{c_y+\rho_y b z}}{(1+\e^{z})^{b}}.
\end{align}

For \eqref{Py.z.identified} to be a valid probabilistic model, we need the following:
\begin{proposition}\label{Prop:Proba.from.Elo}
    The model \eqref{Py.z.identified} may be treated as a probability $\PR{Y_t=y|z}$, if and only if 
    \begin{align}
    \label{b.from.L}
        b&=L-1,\\
    \label{rho.from.y.L}
        \rho_y&=y/(L-1),\\
    \label{a.from.y.L}
        c_y&=\log\nchoosek{L-1}{y}.
    \end{align}
\textbf{Proof:} For \eqref{Py.z.identified} to satisfy the law of total probability, the following must hold:
\begin{align}
    \sum_{y=0}^{L-1}\tilde\mfP_y(z)&=1,\quad \forall z,\\
\label{sum.proba=1.2}
    \sum_{y=0}^{L-1}\e^{c_y + \rho_y b z}&=(1+\e^{z})^{b}.
\end{align}

Since \gls{lhs} of \eqref{sum.proba=1.2} is a polynomial, \gls{rhs} must also be a polynomial of the same order. This requires $b=L-1$, which yields \eqref{b.from.L} and, from a binomial expansion $(1+\e^z)^{L-1}=\sum_{y=0}^{L-1}\nchoosek{L-1}{y}\e^{zy}$, we find \eqref{rho.from.y.L}-\eqref{a.from.y.L}.\hfill$\blacksquare$
\end{proposition}

Thus, only if $\rho_y$ is uniformly distributed in the interval $[0,1]$ as shown in \eqref{rho.from.y.L}, we are able to find the model \eqref{Py.z.identified} for which the \gls{ml}+\gls{sg} ranking \eqref{ML.update} is exactly equivalent to the Elo ranking \eqref{Elo.rating.multilevel}.

We note that a similar result was provided in \cite{Szczecinski20}, where it was shown that, for ternary results ($L=3$, \ie taking into account draws and $\bdelta=[0,0.5,1]$ which satisfies \eqref{rho.from.y.L}), the Elo algorithm \eqref{Elo.rating.multilevel} may be seen as the \gls{ml}+\gls{sg} estimation for the particular form of the Davidson model \cite{Davidson70}. 

However, if the practitioner decides to use $\rho_y \neq y/(L-1)$, we cannot find a model that yields the \emph{exact} equivalence between the algorithm \eqref{Elo.rating.multilevel} and \gls{ml}+\gls{sg} estimation. The Elo algorithm still ``works", \ie calculates the skills $\btheta_t$. But they are solution of the \gls{sg} optimization of the objective function defined in \eqref{ell.from.integration.result}. The Elo algorithm does not optimize the likelihood simply because there is no underlying probabilistic model!

\subsection{Statistician's perspective: Ordinal model and ranking algorithm}\label{Sec:ordinal.models}

The first step for a statistician is to define a model. We use here the \gls{ac} model which was already shown in \cite{Szczecinski22} to be related to the Elo ranking and which relies on a multinomial logistic expression, \cite{Bock72}, \cite{Tutz20}, \cite{Egidi21} 
\begin{align}
\label{AC.model}
\mfP_{y}(z)&= \frac{\e^{\alpha_y + \delta_y z}}{\sum_{l=0}^{L-1} \e^{\alpha_l + \delta_l z}},
\quad y=0,\ld, L-1;
\end{align}
where, to enforce $\mfP_y(z)=\mfP_{L-1-y}(-z)$ given in \eqref{P_y(z).symmetry}, we set
\begin{align}
\label{alpha.symmtry}
    \alpha_y&=\alpha_{L-1-y}\\
\label{delta.symmtry}
    \delta_y&=1-\delta_{L-1-y}.
\end{align}
and, without any loss of generality\footnote{Because the model does not change if we use $\alpha_l\leftarrow \alpha_l - \alpha_\tnr{ref}$, and/or $\delta_l\leftarrow \delta_l - \delta_\tnr{ref}$, \eg with $\alpha_{\tnr{ref}}=\alpha_0$ and $\delta_{\tnr{ref}}=\delta_0$. Furthermore, we can replace $\delta_l\leftarrow\delta_l \beta$, and $z\leftarrow z/\beta$, \eg $\beta=1/\delta_{L-1}$ and the latter becomes an inherent scale.} we may set
\begin{align}
\label{alpha.delta.constraints}
    \alpha_0&=\alpha_{L-1}=0,\quad \delta_0=1-\delta_{L-1}=0.
\end{align}

The coefficients are gathered in the vectors $\bdelta=[0,\delta_1,\ld,\delta_{L-2},1]$ and $\balpha=[0,\alpha_1,\ld,\alpha_{L-2},0]$, where, due to the constraints \eqref{alpha.symmtry}-\eqref{alpha.delta.constraints}, we can freely set $L-2$ parameters. In particular, in the binary outcome, $L=2$, there is no parameter to define because we have $\bdelta=[0,1]$, $\balpha=[0,0]$, and the model \eqref{AC.model} is the logistic model of Sec.~\ref{Sec:Statistician.perspective}. On the other hand, for $L=3$, $\bdelta=[0,0.5,1]$ $\balpha=[0,\alpha_1,0]$, and we have one free parameter: $\alpha_1$.

\begin{figure}
    \centering
    \includegraphics[width=0.9\linewidth]{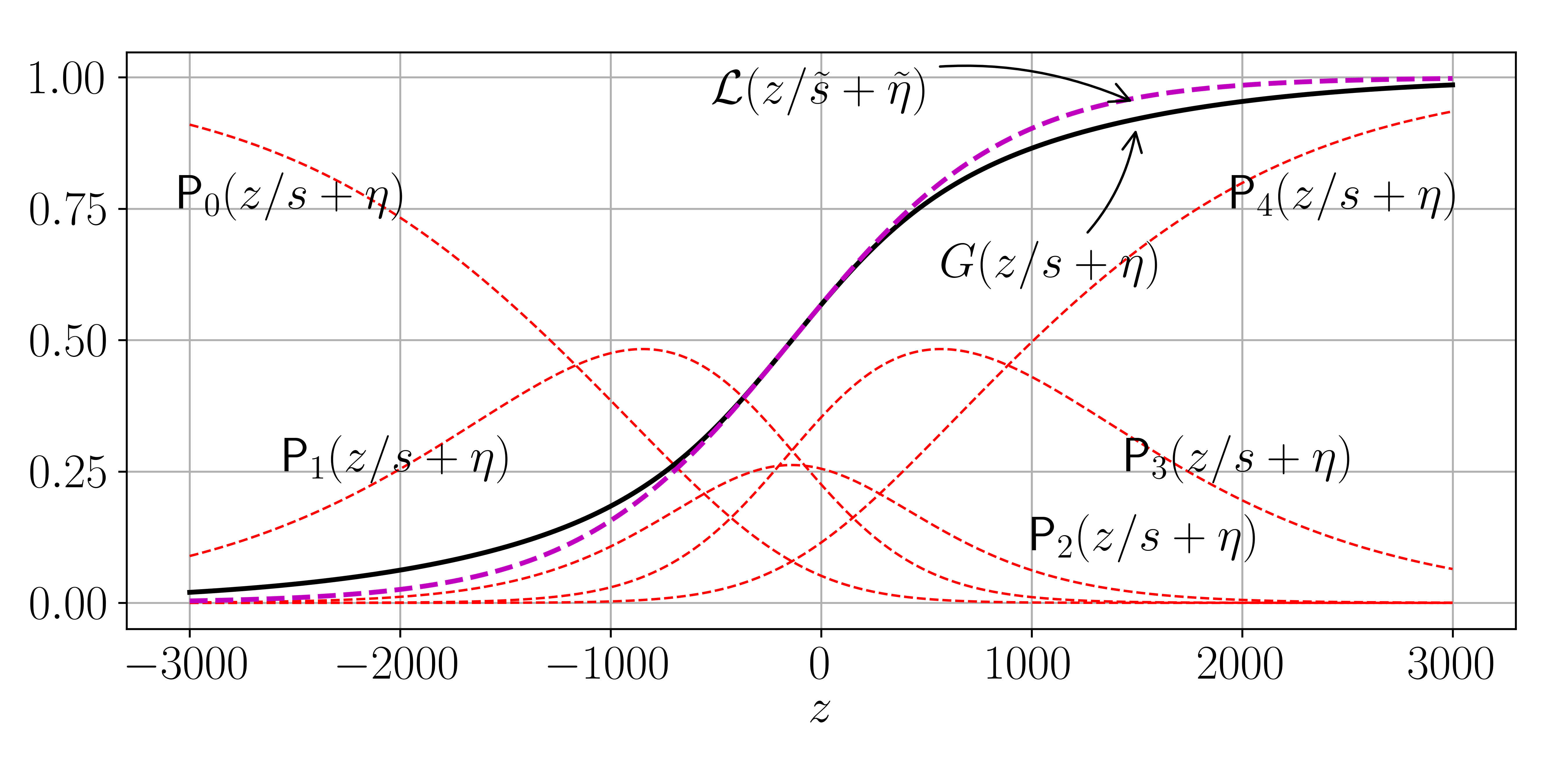}
    \caption{Conditional probability functions $\mfP_y(z/s+\eta)$ \eqref{AC.model} defining the \gls{ac} model with $\balpha$ and $\bdelta$ given in \eqref{alpha.example.L=5} and \eqref{delta.example.L=5}, $s=174$, and $\eta=0.8$. The solid thick line denotes the expected value of the score, $G(z/s+\eta)$, given in \eqref{G(z).AC} and solid dashed line denotes the approximation of the latter using a canonical function $\mcL(z/\tilde{s}+\tilde\eta)$ with $\tilde{s}$ and $\tilde\eta$ in  \eqref{from.s.tilde.to.s} and \eqref{tilde.eta.from.eta}.}
    \label{Fig:Example.AC.L=5}
\end{figure}

Examples of the function $\mfP_{y}(z)$ are shown in Fig.~\ref{Fig:Example.AC.L=5} for $L=5$ and 
\begin{align}
\label{alpha.example.L=5}
    \balpha&=[0,1.3,1.2,1.3,0]\\
\label{delta.example.L=5}
    \bdelta&=[0, 0.22, 0.5, 0.78, 1]
\end{align}
taken from \cite[Table~2]{Szczecinski22}.\footnote{Obtained from the \gls{epl} results by discretizing the goal difference $g$ into $l=5$ events: $g\le -3$ (strong loss), $g\in\set{-2,-1}$ (weak loss), $g=0$ (draw), $g\in\set{1,2}$ (weak win), and $g\ge 3$ (strong win), where the coefficients are calculated from the frequency of these events. Note that we ignore \gls{hfa}.

Moreover, because \cite{Szczecinski22} uses $10^{\alpha_y}$ which is the same as $\e^{\alpha_y\log 10 }$, the coefficients $\alpha_y$ we show here are obtained by multiplying those in \cite[Table~2]{Szczecinski22} by a factor $\beta_{\e\rightarrow 10}=\log 10\approx 2.3$ yielding $\alpha_1=0.51$, $\alpha_2=0.57$ and $\eta=0.4$. No changes are required in $\delta_y$ because, as we discussed at the end of Sec.~\ref{Sec:Elo.algorithm}, the change in the exponent's basis will be absorbed by the scale.
}

\textbf{ML+SG Ranking}

To obtain the \gls{ml}+\gls{sg} ranking \eqref{ML.update.scale.HFA} we apply \eqref{dot.ell.definition} to \eqref{AC.model}
\begin{align}
\label{dot.ell_y(z).AC}
    \dot\ell_y(z)&=\delta_y - G^\tnr{AC}(z)\\
\label{G(z).AC}
    G^\tnr{AC}(z) & =\frac{\sum_{l=1}^{L-1}\delta_l\e^{\alpha_l + \delta_l z}}{\sum_{l=0}^{L-1} \e^{\alpha_l + \delta_l z}},
\end{align}
which used in \eqref{ML.update.scale.HFA} yields
\begin{align}
    \label{ML+SG.AC.model}
    \btheta_{t+1} 
    &= \btheta_{t} + \mu s \bx_t [\delta_{y_t} - G^\tnr{AC}(z_t/s+\eta h_t)].
\end{align}

We note that \eqref{ML+SG.AC.model} is a \gls{gelo} algorithm proposed in \cite{Szczecinski22}.

The obvious relationship to \eqref{Elo.rating.multilevel} cannot be missed if we set $\rho_y=\delta_y$ and treat $G^\tnr{AC}(z_t)$ as ``the expected score''; in fact, it \emph{is} the mathematical expectation of the score $\delta_{y_t}$, \ie
\begin{align}
\label{G(z_t/s)}
    \Ex_{Y_t|z_t}[\delta_{Y_t}]&=\sum_{y=0}^{L-1}\delta_y\mfP_{y}(z_t)=G^\tnr{AC}(z_t).
\end{align}
where the last equality is obtained by using \eqref{AC.model} and \eqref{G(z).AC}.

As expected, for $L=2$, we recover the binary-outcome model and all formulas are consistent with those shown in Sec.~\ref{Sec:Statistician.perspective}. 

However, for $L>2$ there is still one difference with the Elo algorithm \eqref{Elo.rating.multilevel}: the form of the expected score \eqref{G(z).AC} is not the same as the logistic function in the latter, and to obtain the full equivalency, we need the following:
\begin{proposition}
    \label{Prop:AC=Elo}
    If $\alpha_y = \log\nchoosek{L-1}{y}$ and $\delta_y= \frac{y}{L-1}$, the following holds:
    \begin{align}\label{G(z)=L(z/tilde.s)}
        G^\tnr{AC}(z/s) &= \mcL(z/\tilde{s}),\\
        \label{tilde.s=s*beta_AC.to.L}
        \tilde{s}&=s\beta_{\tnr{AC}\rightarrow\mcL},
    \end{align}
    where  
    \begin{align}\label{beta.AC.to.L.Proposition}
        \beta_{\tnr{AC}\rightarrow\mcL}&=L-1
    \end{align}
    is the scale adjustment factor allowing us to replace the expected score in the \gls{ac} model \eqref{G(z).AC} with the canonical logistic function $\mcL(\cd)$.
\\
\textbf{Proof}: Appendix~\ref{Sec:Proof.Lemma.AC}.
\end{proposition}

Thus, for the matches with $L$ outcomes, if we set $\delta_y=y/(L-1)$, the Elo algorithm aims at fitting the skills $\btheta_t$ to observations $y_t$, using the \gls{ac} model with parameters $\alpha_y = \log\nchoosek{L-1}{y}$. This property of the \gls{ac} model motivated \cite{Szczecinski22} to consider the \gls{gelo} algorithm \eqref{ML+SG.AC.model} to be a ``natural'' generalization of the Elo algorithm. This does not mean that these values $\alpha_y$ are optimal in any way as we did not make any effort to identify them from the data.

On the other hand, if we violate the condition of Proposition~\ref{Prop:AC=Elo}, $\alpha_y\neq \log\nchoosek{L-1}{y}$ or $\delta_y\neq y/(L-1)$, there is no Elo-algorithm which will provide the same results as the \gls{gelo} algorithm.

This modeling mismatch is another layer of approximation on top of the approximate optimization characteristic of the \gls{sg} approach. To see if it may be acceptable with respect to the \gls{ac} model, we may compare $G^\tnr{AC}(z/s)$ to the ``equivalent'' canonical form $\mcL(z/\tilde{s})$, where, motivated by the scale adjustment required in \eqref{tilde.s=s*beta_AC.to.L}, we can conveniently adjust $\tilde{s}$.


\subsubsection{Approximate equivalence of the expected score}\label{Sec:approximate.equivalent}
Let us explore the idea that $\mcL(z/\tilde{s})$ can approximate the expected score $G^\tnr{AC}(z/s)$, for any $\balpha$ and $\bdelta$ and not only those specified in Proposition~\ref{Prop:AC=Elo}.

By analogy to the approach already used in Sec.~\ref{Sec:Scaling.HFA}, for the approximate equivalence between the Elo ranking and the \gls{gelo} ranking (for the \gls{ac} model), we require the algorithmic update for $z=0$ to be identical, \ie
\begin{align}
\label{gradient.equivalence}
    [\rho_y - \mcL(z/\tilde{s})]_{z=0} 
    &= [\delta_y-G^\tnr{AC}(z/s)]_{z=0},
\end{align}
which is satisfied if $\delta_y=\rho_y$, 
and we postulate the equivalence of their first derivatives 
\begin{align}    
\label{2nd.derivative.equivalence}
    \frac{1}{\tilde{s}}\dot\mcL(z/\tilde{s})|_{z=0}
    &=\frac{1}{s}\frac{\dd}{\dd z}{G}^\tnr{AC}(z/s)|_{z=0}.
\end{align}

The condition \eqref{2nd.derivative.equivalence} is expressed as
\begin{align}
\label{equivalence.derivatives.AC}
    \frac{1}{4\tilde{s}}
    &=
    \frac{1}{s}\left(
    \frac{\big(\sum_{l=0}^{L-1}\delta_l^2\e^{\alpha_l + \delta_l z/s}\big)}
    {\big(\sum_{l=0}^{L-1} \e^{\alpha_l + \delta_l z/s}\big)}
    -
    \frac{\big(\sum_{l=0}^{L-1}\delta_l\e^{\alpha_l + \delta_l z/s}\big)^2}
    {\big(\sum_{l=0}^{L-1} \e^{\alpha_l + \delta_l z/s}\big)^2}
    \right)\Big|_{z=0},
\end{align} 
which yields 
\begin{align}
\label{from.s.tilde.to.s}
    \tilde{s}&=s\beta_{\tnr{AC}\rightarrow\mcL},\\
\label{beta.from.AC}
    \beta_{\tnr{AC}\rightarrow\mcL}&=
    \Big[4
    \frac{\big(\sum_{l=0}^{L-1}\delta_l^2\e^{\alpha_l}\big)}
    {\big(\sum_{l=0}^{L-1} \e^{\alpha_l}\big)}
    -
    1\Big]^{-1}.
\end{align}

To illustrate how this works for $L=5$, we have shown the approximation $\mcL(z/\tilde{s}+\tilde\eta)$ in Fig.~\ref{Fig:Example.AC.L=5}, while the common case of the three-level matches, \ie $L=3$ is discussed in the following example.

\begin{example}[$L=3$]\label{Ex:expected.score.L=3}
In games with three-level outcomes, \ie when $L=3$, $\bdelta=[0,0.5,1]$ and $\balpha=[0,\alpha_1,0]$, we have
\begin{align}
\label{beta.from.kappa_1}
    \beta_{\tnr{AC}\rightarrow\mcL} &= 1+0.5\e^{\alpha_1},
\end{align}
and the approximation results are shown in Fig.~\ref{fig:mcL.vs.G.kappa.L=3} for different values of $\alpha_1$. From Proposition~\ref{Prop:AC=Elo} we know that for $\alpha_1\approx 0.7$, we have $\beta_{\tnr{AC}\rightarrow\mcL}=2$ and $G(z/s)=\mcL(z/\tilde{s})$, \ie there is no approximation error.

While Fig.~\ref{fig:mcL.vs.G.kappa.L=3} provides us with a cautionary note on the model mismatch, it seems to be important only for
large $\alpha_1$, say $\alpha_1\ge 0.7$. Thus, the question is whether such models are suitable to describe the results of the matches? 

This can be answered using data, but a glimpse of an insight may be obtained assuming that all skills are comparable and thus $z_t\approx 0$ (as may happen, with small $v_\theta$, \eg in the leagues competitions), so we can find the probability of the draw as (see also \cite[Eq.~(47)]{Szczecinski20})
\begin{align}\label{Pr.draw}
    \mfP_1(z)|_{z\approx 0} \approx \frac{\e^{\alpha_1}}{2+\e^{\alpha_1}}
\end{align}

Then, for $\alpha_1= -1.0,0.0, 0.7, 1, 2$, which we used in Fig.~\ref{fig:mcL.vs.G.kappa.L=3}, we obtain, respectively, the probability of the draw \eqref{Pr.draw} $\approx 0.16, 0.33, 0.5, 0.58, 0.79$.

The probability of the draw larger than 0.5 (\ie for $\alpha_1>0.7$, where the mismatch is apparent in Fig.~\ref{fig:mcL.vs.G.kappa.L=3}) seems rather unlikely in popular sports with three level outcomes, such a football, basketball. Then, we would rather use $\alpha_1<0.7$ and the Elo algorithm's $\mcL(z/\tilde{s})$ is a reasonable approximation of the expected score $G^\tnr{AC}(z/s)$. In competition where large $\alpha_1$ makes sense (\eg in chess, the probability of draws may be large) the modeling mismatch will be more important.

\begin{figure}
    \centering
    \includegraphics[width=0.9\linewidth]{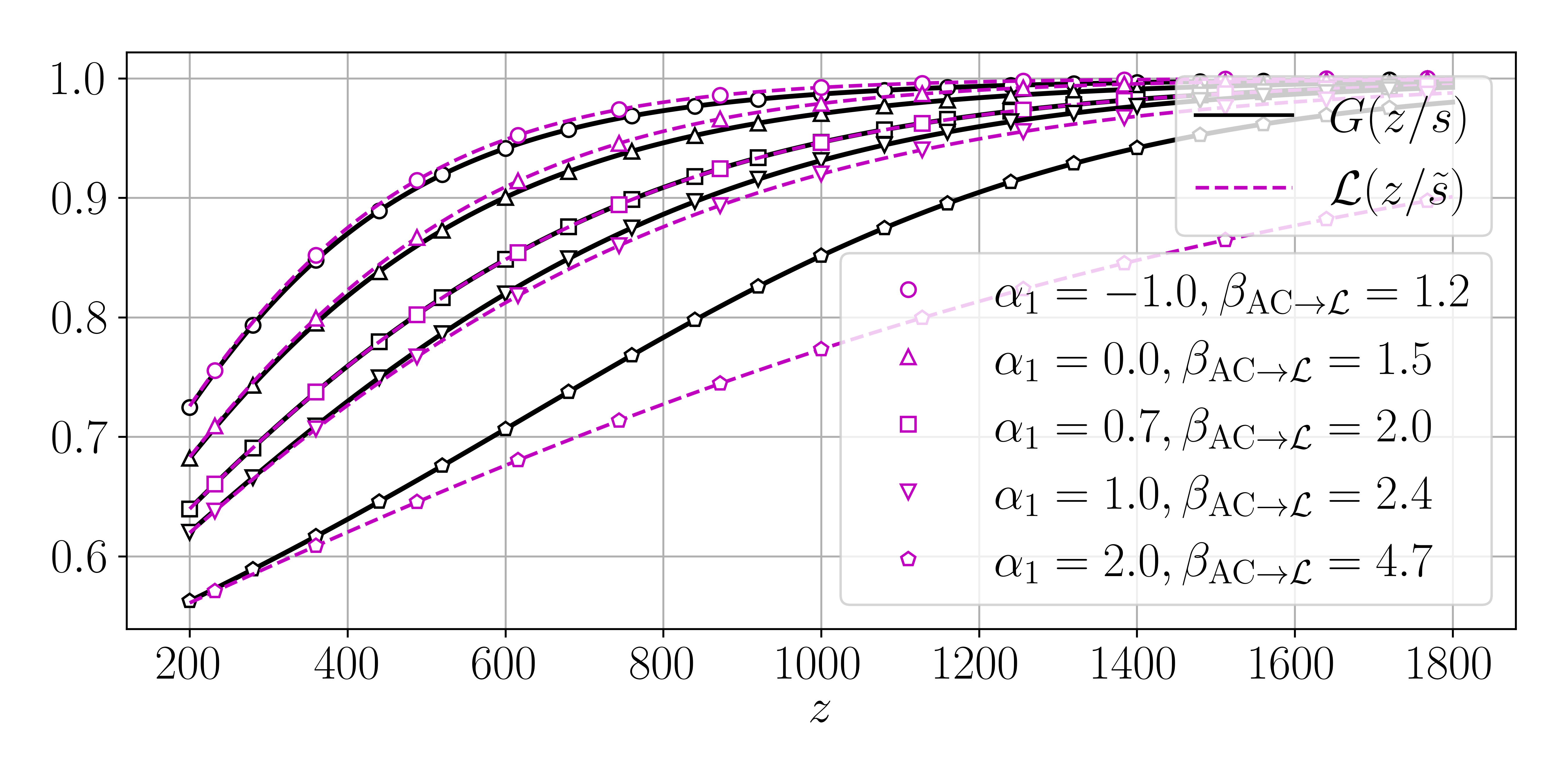}
    \caption{Comparison between $G(z/s)$ and its approximation $\mcL(z/\tilde{s})$, for $L=3$, $\tilde{s}=s\beta_{\tnr{AC}\rightarrow\mcL}$, $\beta_{\tnr{AC}\rightarrow\mcL}$ given in \eqref{beta.from.kappa_1}, $\bdelta=[0,0.5,1]$, $\balpha=[0,\alpha_1,0]$, where the values of $\alpha_1$ are given in the legend; $s= 174$. For smaller values of $z$, the curves practically superimpose. For $\alpha_1=\log 2\approx 0.7$, we have a true equivalence of the expected scores, \ie $G(z/s)=\mcL(z/\tilde{s})$, where $\tilde{s}=2s$.}
    \label{fig:mcL.vs.G.kappa.L=3}
\end{figure}
\end{example}

\subsection{Ranking and prediction: decoupling the models}\label{Sec:model.decoupling}

We want to combine the practitioner's and statistician's perspectives: the goal is not to change the Elo algorithm but to provide statistical tools to evaluate its  performance and/or interpret its results.

The core idea is that the ranking is not necessarily based on explicit probabilistic model but provides results which are close to those that we might obtain with a full \gls{ac}, unknown to us, model. Finding this ``implicit'' model is the problem we want to address. That is, we allow ourselves to decouple the modelling used in the fit (ranking) from the modelling in the prediction. This concept should not be surprising because we already introduced this idea when changing the scale of the model in the prediction to take into account estimation errors.

Motivated by analysis which follows \eqref{gradient.equivalence}, we assume that $\btheta_t$ produced by the Elo ranking with scores $\brho$ and the scale $s$, is an approximate \gls{ml} solution based on the \gls{ac} model. This is not to say that the \gls{ac} model is optimal in any sense, but it is a plausible candidate suggested by the similarity of the Elo algorithms \eqref{Elo.rating.multilevel} and the \gls{gelo} algorithm \eqref{ML+SG.AC.model} defined for the \gls{ac} model.

We naturally assume that the parameters of the \gls{ac} model are given by $\bdelta=\brho$ (again, without any pretense of optimality) and we need to find the parameters $\balpha$, as well as the corresponding scale $\hat{s}$ which can be used in the performance evaluation.

Why shouldn't we use $\balpha$ specified in Proposition~\ref{Prop:AC=Elo}? First, because it applies only for $\delta_y=y/(L-1)$. However, even if we worked with such values (\eg normally used when $L=3$, \ie in win/draw/loss games) we have no guarantee that the model with $\alpha_y=\log\nchoosek{L-1}{y}$ given in Proposition~\ref{Prop:AC=Elo} is the best to describe the data. The motivation for decoupling of the models should be obvious: if the model specified by Proposition~\ref{Prop:AC=Elo} is mismatched with respect to the data, it would be foolish to use the same model for prediction!

Our objective is thus twofold: using the data $y_t$ and estimated skills $\btheta_t$ if necessary, we will find (i) the parameters $\hat\balpha$ and $\hat\eta$ of the \gls{ac} model and (ii) the scale $\hat{s}=s\beta$ that will be used in the prediction $\log\PR{Y_t=y|\btheta_t}=\ell_y\big((\bx\T_t\btheta_t/\hat{s}+\hat\eta;\hat\balpha\big)$, where $\btheta_t$ are obtained from the Elo algorithm, and $\ell_y(z;\balpha)=\log\mfP_y(z;\balpha)$ given in \eqref{AC.model} makes the dependence on $\balpha$ explicit.

The re-scaling $\hat{s}=s\beta$ plays now a double role. First, it is required to align the expected score of the \gls{ac} model (we use for prediction) with the one used in the Elo ranking: remember we want to ensure $\mcL(z/s)\approx G^\tnr{AC}(z/(s\beta_{\mcL\rightarrow\tnr{AC}}))$, where $\beta_{\mcL\rightarrow\tnr{AC}}=1/\beta_{\tnr{AC}\rightarrow\mcL}$. Second, through $\beta_\tnr{err}$, it takes into account the estimation errors in $\btheta_t$ akin to \eqref{tilde.beta.binary}.\footnote{This may seem somewhat ad hoc, but the convolution with the Gaussian in Appendix~\ref{Proof.empirical.expected.score} has the effecting of spreading the functions, which is well modeled as a change of scale.} This compound effect can be written as
\begin{align}\label{hat.beta=tilde.beta.beta.AC.L}
    \beta&\approx\beta_\tnr{err}\beta_{\mcL\rightarrow\tnr{AC}}
    =\frac{\beta_\tnr{err}}{\beta_{\tnr{AC}\rightarrow\mcL}}.
\end{align}

Since $\beta_\tnr{err}$ and $\beta_{\tnr{AC}\rightarrow\mcL}$ are entangled in \eqref{hat.beta=tilde.beta.beta.AC.L}, it is simpler to identify $\beta$ directly from the data together with the parameters $\balpha$ and $\eta$ of the \gls{ac} model, \ie 
\begin{align}\label{hat.alpha.scale.batch}
    \hat\balpha, \hat\eta, \hat\gamma &= \argmax_{\balpha, \eta, \beta} \sum_{t\in\mcT^\tnr{train}}\ell_{y_t}\Big(\gamma z_t/s+\eta h_t; \balpha\Big),\\
    \hat\beta & = 1/\hat\gamma
\end{align}
where $z_t=\bx\T_t\btheta_t$ is calculated from the skills $\btheta_t$, and $\mcT^\tnr{train}$ is the set of indices to the training data that should be different from the one used for testing. This is the pseudo-model identification akin to \eqref{hat.s.from.data}. Again, we use $\gamma=1/\beta$ to guarantee the optimization uniqueness under concavity in $\gamma$, and we added the \gls{hfa} indicator, $h_t$ for completeness.

We can now use the log-score formula \eqref{neg.log.scores.definition.corrected} adapted to the \gls{ac} model:
\begin{align}\label{neg.log.scores.adaptive}
    \mf{LS} = -\frac{1}{|\mcT^\tnr{test}|}\sum_{t\in\mcT^\tnr{test}} \ell_{y_t}\left(\frac{z_t}{s\hat\beta}+\hat\eta h_t;\hat\balpha\right),
\end{align}
where $\mcT^\tnr{test}$ contains indices of matches played \emph{after} those in the training set.\footnote{Then, $\hat\balpha$, $\hat\eta$ and $\hat\beta$, do not depend on $(y_t, z_t) \quad t\in\mcT^\tnr{test}$} 

The decoupling should now be clear: The Elo algorithm~\eqref{Elo.rating.multilevel} fits the skills $\btheta_t$ to observations via optimization that may, but need not, correspond to the \gls{ac} model (with $\bdelta=\brho$ and,  if $
\rho_l=l/(L-1)$, $\balpha$ specified in  Proposition~\ref{Prop:AC=Elo}). The prediction model, on the other hand, is parameterized by  $\hat{\balpha}$, $\hat{\eta}$, $\hat{\beta}$, which are estimated from outcomes $y_t$ and from $z_t$ produced by the ranking algorithm. Only $\bdelta=\brho$ is shared between the models.

This decoupling principle already appeared in a simple form in Sec.~\ref{Sec:Performance evaluation}, where the prediction scale $\hat{s}$ and the \gls{hfa} $\hat\eta$ differ from the nominal $s$ and $\eta$; in the multilevel setting, the decoupling is conceptually less obvious because the underlying model is only implicit and/or approximate.

\subsubsection{Simplifications: avoid optimization}\label{Sec:Simplifications}
Instead of a potentially tedious optimization, the problem simplifies if we assume that $z_t$ are sufficiently small to use $z_t\approx 0$. Then, \eqref{hat.alpha.scale.batch} may be written 
\begin{align}\label{hat.alpha.scale.batch.z=0}
    \hat\balpha, \hat\eta &= \argmax_{\balpha, \eta} 
    (1-\ov\mfP_\tnr{hfa})\sum_{y\in\mcY}\ov\mfP_{y,\tnr{neut}} \ell_y(0)+ \ov\mfP_\tnr{hfa}\sum_{y\in\mcY}\ov\mfP_{y,\tnr{hfa}} \ell_y(\eta),
\end{align}
where $\ov\mfP_{y,\tnr{neut}}$ and $\ov\mfP_{y,\tnr{hfa}}$ are frequencies, observed in the training set $\mcT_\tnr{train}$, of outcomes $y$ when $h_t=0$ (neutral venue) or when $h_t$ (\gls{hfa} effect); $\ov\mfP_\tnr{hfa}$ is the overall frequency of playing on the home venues. This is the value of the simplification we can gather all terms with the same/similar values of $z_t\approx 0$.

Even with the simplifcation, in general, there is no closed-form solution to this problem. However, when the matches are venue-homogenous (always played on neutral or always with \gls{hfa}, \ie $\ov\mfP_\tnr{hfa}=1$ or $\ov\mfP_\tnr{hfa}=0$), we have \cite[Sec.~3.2]{Szczecinski22}
\begin{align}
\label{alpha.1.initialization}
    \hat\alpha_y & = 0.5\log\left(\frac{\ov\mfP_y \ov\mfP_{L-1-y}}{\ov\mfP_0 \ov\mfP_{L-1}}\right);
\end{align}
here, $\mfP_y$ should denote the frequency of events (on neutral on \gls{hfa} venues) but, if we deal with mixed venues, \ie $0<\ov\mfP_\tnr{hfa}<1$, in practice, we may use the venue-blind frequencies of events $\ov\mfP_y$.\footnote{We may want to first symmetrize the neutral venue frequencies, $\ov\mfP_{y,\tnr{neut}}\leftarrow 0.5(\ov\mfP_{y,\tnr{neut}}+\ov\mfP_{L-1-y,\tnr{neut}})$, and then reweight all results $\ov\mfP_{y}=(1-\ov\mfP_\tnr{hfa})\ov\mfP_{y,\tnr{neut}}+\ov\mfP_\tnr{hfa}\ov\mfP_{y,\tnr{hfa}}$.}

The \gls{hfa} can only be determined from the matches when $h_t=1$. Zeroing the derivative of the function under maximization in \eqref{hat.alpha.scale.batch.z=0}, with respect to $\eta$ yields
\begin{align}
\label{FOC:eta}
    \frac{\dd}{\dd \eta}(\cd)&=
    \sum_{y\in\mcY}\ov\mfP_{y,\tnr{hfa}}\Big(\delta_y-\sum_{l\in\mcY}\delta_l\mfP_l(\eta)\Big)
    =\ov\delta_\tnr{hfa} - G^\tnr{AC}(\eta)|_{\eta=\hat\eta}=0,\\
\label{approx.eta}
    \ov\delta_\tnr{hfa}&=G^\tnr{AC}(\hat\eta)\approx \mcL(\hat\eta/\beta_{\tnr{AC}\rightarrow \mcL}),
\end{align} 
where $\ov\delta_\tnr{hfa}=\sum_{y\in\mcY}\ov\mfP_{y,\tnr{hfa}}\delta_y$ is the score for \gls{hfa} averaged over matches in $\mcT_\tnr{train}$, and 
\eqref{approx.eta} exploits approximate equivalence $G^\tnr{AC}(z)\approx\mcL(z/\beta_{\tnr{AC}\rightarrow \mcL})$ postulated in  \eqref{gradient.equivalence}-\eqref{beta.from.AC} with $\beta_{\tnr{AC}\rightarrow \mcL}$ given in \eqref{beta.from.AC}. 

We can solve \eqref{approx.eta} as
\begin{align}
    \label{eta.0.initialization}
    \hat\eta &\approx \mcL^{-1}(\ov\delta)\beta_{\tnr{AC}\rightarrow \mcL},
\end{align}
where $\mcL^{-1}(\cd)$ is the logit function and we note that \eqref{eta.0.initialization} is different from \cite[Eq.(44)]{Szczecinski22} because we deal with $\bdelta=\brho$, which are not (necessarily) optimized to fit $\ov\mfP_y$.

Further, ignoring the estimation noise in \eqref{hat.beta=tilde.beta.beta.AC.L} (\ie  setting $\beta_\tnr{err}=1$), we obtain
\begin{align}\label{beta.0.initialization}
    \hat\beta = 1/\beta_{\tnr{AC}\rightarrow \mcL}.
\end{align}

The closed-form formulas \eqref{alpha.1.initialization}, \eqref{eta.0.initialization}, and \eqref{beta.0.initialization} can be used directly in \eqref{neg.log.scores.adaptive} (with $z_t$ obtain from the Elo ranking).

\begin{example}[Pseudo-model identification]\label{Ex:find_alpha_and_scale}
For $\btheta^*$ we used in Example~\ref{Example:Illustration.convergence} we generate outcomes of the matches using the \gls{ac} model \eqref{AC.model} with parameters $\eta^*=0.35$ and $\alpha_1^*=-0.4$. We use $\bdelta=[0, 0.5, 1]$, the Elo algorithm \eqref{Elo.practitioner.compact.scale.2} with scale $s=174$, $\eta=0$, and steps $K=20$ and $K=60$. 

We define the set $\mcT^\tnr{train}=\set{4000, \ld, 7999}$ for pseudo-model identification in \eqref{hat.alpha.scale.batch}, and a testing set $\mcT^\tnr{test}=\set{8000, \ld, 12000}$ for evaluation in \eqref{neg.log.scores.adaptive}. The matches indexed with $t<4000$ are only used to  ensure convergence. The frequency $\ov{\mfP}_y$  needed in \eqref{alpha.1.initialization}--\eqref{beta.0.initialization} is estimated from outcomes $y_t, t\in\mcT^\tnr{train}$.


To assess the benefit of the full adaptation \eqref{hat.alpha.scale.batch} relative to non-adaptive methods, we consider the following cases, ordered by increasing use of the data and/or prior information:
\begin{itemize}
    \item \textbf{Conventional}: Based on the premise that the model used in ranking \emph{must} be used for prediction, we use $\hat\eta=0$, 
    $\hat\alpha_1 = \log 2\approx 0.7$, and $\hat\beta = 1/\beta_{\tnr{AC}\rightarrow\mcL}=0.5$, as implied by Proposition~\ref{Prop:AC=Elo}.  this illustrates the ``coupling" of the models which ignores model mismatch (as well as the estimation noise); it is a ``straw man'' baseline, also used in \cite{Szczecinski22a}.

    \item \textbf{Simple, no HFA}:
    $\hat\alpha_1$ and $\hat\beta$ are found from the frequencies of outcomes via \eqref{alpha.1.initialization},\eqref{beta.0.initialization} and we set $\hat\eta = 0$. This is done to verify how much we lose by ignoring the \gls{hfa} in the performance evaluation.

    \item \textbf{Simple, with HFA}: as above, and $\hat\eta$ is calculated from \eqref{eta.0.initialization}. 
    
    \item \textbf{Simple, optimal scaling}:
    $\hat\beta$ is found solving \eqref{hat.alpha.scale.batch} under constraints of fixed $\hat\alpha_1$ and $\hat\eta$ obtained from \eqref{alpha.1.initialization}-\eqref{eta.0.initialization}. 

    \item \textbf{Fully adaptive}:
    $\hat\balpha$, $\hat\eta$ and $\hat\beta$ are found from \eqref{hat.alpha.scale.batch}.

    \item \textbf{\gls{gelo}}: The results of the \gls{gelo} algorithm \eqref{ML+SG.AC.model} with values $\alpha^*_1$ and $
    \eta^*$ are used to calculate the log-score \eqref{neg.log.scores.adaptive} with $\hat\eta=\eta^*$ and $\hat\alpha_{1}=\alpha_1^*$. Due to the perfect alignment of the models used in data generation, ranking, and prediction,  $\hat\beta>1$ appears only to correct for the presence of estimation noise, \ie $\hat\beta\approx\beta_\tnr{err}$.
\end{itemize}
The lower bound for the log-score \eqref{neg.log.scores.adaptive} is obtained with the ground truth, \ie by replacing $z_t$ with $z^*_t=\bx\T_t\btheta^*$ and setting $s=1$, $\beta\equiv 1$, $\hat\eta=\eta^*$, and $\hat\alpha_{1}=\alpha_1^*$; this yields the conditional entropy of $Y_t|z^*_t$ averaged over $z^*_t$, which no data-driven method can surpass.

In Table~\ref{tab:example4} we compare the log-scores \eqref{neg.log.scores.adaptive} of the methods listed above against the ground-truth lower bound. When $\hat\alpha_1$, $\hat\eta$, $\hat\beta$ are averaged (over 200 realizations) of data, we show, in parentheses (mean, standard deviation). 

\begin{table}[t]
\centering
\caption{Log-score \eqref{neg.log.scores.adaptive}, $\hat{\alpha}_1$, $\hat{\beta}$, and $\hat{\eta}$
for ternary matches with true parameters $\alpha^*_1 = -0.4$, $\eta^* = 0.35$. Elo runs with the scale $s = 174$, and two values of $K$. The mean/standard deviation are shown before/after $\pm$. Lower values of $\mf{LS}$ indicate better prediction. When obtained from the data, we show (mean, std) obtained from $J=200$ realizations; numbers not in brackets indicate predefined values.}
\label{tab:example4}
\resizebox{\linewidth}{!}{
\begin{tabular}{lcccc}
\toprule
Method & $\hat{\alpha}_1$ & $\hat{\beta}$ & $\hat{\eta}$ & $\mathsf{LS}$ \\
\midrule
\multicolumn{5}{l}{$K = 20$} \\
\midrule
Conventional                  & $0.7$                     & $1/2$                          & $0$                             & $(1.118, 0.010)$ \\
Simple, no HFA                & \multirow{3}{*}{$(-0.53, 0.04)$} & \multirow{2}{*}{$(0.77, 0.01)$} & $0$                             & $(0.999, 0.007)$ \\
Simple, with HFA              &                                   &                                 & \multirow{2}{*}{$(0.29, 0.04)$} & $(0.990, 0.008)$ \\
Simple, optimal scaling       &                                   & $(0.89, 0.02)$                  &                                 & $(0.989, 0.008)$ \\
Fully adaptive                & $(-0.42, 0.04)$                   & $(0.86, 0.02)$                  & $(0.34, 0.04)$                  & $(0.987, 0.007)$ \\
G-Elo                          & $\alpha^*_1$                           & $(1.12, 0.03)$                  & $\eta^*$                         & $(0.984, 0.007)$ \\
\midrule
\multicolumn{5}{l}{$K = 60$} \\
\midrule
Conventional                  & $0.7$                     & $1/2$                          & $0$                             & $(1.161, 0.011)$ \\
Simple, no HFA                & \multirow{3}{*}{$(-0.53, 0.04)$} & \multirow{2}{*}{$(0.77, 0.01)$} & $0$                             & $(1.026, 0.008)$ \\
Simple, with HFA              &                                   &                                 & \multirow{2}{*}{$(0.29, 0.04)$} & $(1.017, 0.008)$ \\
Simple, optimal scaling       &                                   & $(1.18, 0.03)$                  &                                 & $(1.004, 0.007)$ \\
Fully adaptive                & $(-0.44, 0.04)$                   & $(1.16, 0.04)$                  & $(0.33, 0.04)$                  & $(1.003, 0.007)$ \\
G-Elo                          & $\alpha^*_1$                           & $(1.40, 0.03)$                  & $\eta^*$                         & $(0.997, 0.007)$ \\
\midrule
Ground truth (from $z^*_t$)   & $\alpha_1^*$                     & $1$                            & $\eta^*$                       & $(0.976, 0.007)$ \\
\bottomrule
\end{tabular}
}
\end{table}

We conclude that:
\begin{enumerate}
    \item \textbf{Model decoupling pays off}: the conventional (coupled) baseline yields the worst log-score by far ($\mf{LS}=1.118$/$1.161$ for $K=20$/$60$), confirming that blindly reusing
    the ranking model for prediction is harmful regardless of the estimation noise level.
 
    \item \textbf{HFA recovery yields a small gain}: Considering $\hat\eta$, yields a consistent log-score gain ($0.01$). For small estimation noise ($K=20$) we practically recover the ground truth performance despite the Elo algorithm ignoring the \gls{hfa}!
 
    \item \textbf{$\hat{\beta}$ is the most important parameter}: it increases from $0.885$ (at $K=20$) to $1.179$ (for $K=60$), reflecting the inflation of $\beta_\tnr{err}$ in \eqref{hat.beta=tilde.beta.beta.AC.L} (which is consistent with the binary case). In fact, with $\hat\alpha_1$ and $\hat\eta$ found from \eqref{alpha.1.initialization} and \eqref{eta.0.initialization}, it is enough to adapt $\hat\beta$ from the data, or---for sufficiently small $K$, use \eqref{beta.0.initialization}. The fully adaptive approach is not necessary! 
    
    \item \textbf{Exact model is not necessary in ranking}: The log-score $\mf{LS}$ based on the Elo algorithm is within one standard deviation from the performance of the \gls{gelo} algorithm (with the optimal scaling necessary to take care of the estimation noise). This suggests that the gains shown in \cite{Szczecinski22} and \cite{Szczecinski22a} were mostly due to the use of the \gls{ac} model for the performance evaluation, and much less due to the ranking algorithm itself!
\end{enumerate}

\end{example}

\subsubsection{On-line adaptation}\label{Sec:Online.adaptation}
Instead of \eqref{hat.alpha.scale.batch}, and more in the on-line spirit of the Elo ranking, we may solve the problem using an on-line mini-batch approach, \eg for $\beta_t=1/\gamma_t$:
\begin{align}\label{hat.beta.mini.batch.update}
    \gamma_{t+1} 
    &= \gamma_{t}+
    \frac{\mu_\gamma}{|\mcT_t|} \sum_{\tau\in\mcT_t}\frac{\dd}{\dd \gamma}\ell_{y_\tau}\!\left(\gamma z_\tau/s+\hat\eta; \hat\balpha\right)\bigg|_{\gamma=\gamma_{t}}\\
    &=\gamma_t + 
    \frac{\mu_\gamma}{|\mcT_t|}\sum_{\tau\in\mcT_t}
    \frac{z_\tau}{s}\left(\delta_{y_\tau} - G^{\tnr{AC}}\big(\gamma_t z_\tau/s+\hat\eta;\hat\balpha\big)\right)
\end{align}
where $\mu_\gamma$ is the adaptation step, and $\mcT_t=\set{t-W+1,\ld, t}$ is the length-$W$ mini-batch of indices.\footnote{With $\mcT_t=\set{t}$ we recover a conventional \gls{sg} approach.} which should be chosen to average out the randomness in the gradients but preserve adaptability; for long competitions it may be at the order of hundreds. Note that we use $\hat\balpha$ and $\hat\eta$ obtained via \eqref{alpha.1.initialization} and \eqref{eta.0.initialization}, which we have already found to perform well. Their on-line adaptation is possible, but we have not found it useful.

\begin{example}[FIFA Men's ranking]
Let us consider $T=5719$ \gls{fifa} international Men's teams matches played from 2018-06-04 till 2024-07-14, recovered from \cite{football_rankings} with given $y_t$ and $\theta_{t,m}$. The \gls{fifa} Elo-style algorithm has expected score $\mcL(z/s;10)$ with $s=600$, which corresponds to using $\mcL(z/\tilde{s})$ with $\tilde{s}=s\beta_{10\rightarrow\e}=s/{\log 10}\approx 260.5$. The \gls{hfa} indicators $h_t$ are found analyzing the website \cite{soccerway} and using AI tools \cite{claude2026}.

We do not need to know \emph{how} the results and/or skills were obtained (penalties, overtime, forfeit, initialization), which is in line with our model decoupling strategy: what matters is how to fit the pseudo-model for performance identification.

We set $\mcT^\tnr{train}=\set{T',\ld,T''-1}$, $\mcT^\tnr{test}=\set{T'',\ld,T}$, with $T'=2000$, $T''=4000$. The values $\hat\alpha_1$, $\hat\eta$, and $\hat\beta$ as well as the log-score \eqref{neg.log.scores.adaptive} are shown in Table~\ref{tab:FIFA}. For scaling obtained via \eqref{hat.beta.mini.batch.update}, $\hat\beta$ is calculated as a mean of $\beta_t$ over $\mcT_\tnr{test}$.

\begin{table}[t]
\centering
\caption{Log-score~\eqref{neg.log.scores.adaptive}, $\hat\alpha_1$, $\hat\beta$, and $\hat\eta$
for \gls{fifa} Men's international matches ($T=5719$ games,
$2018\text{-}06\text{-}04 \to 2024\text{-}07\text{-}14$);
$\mcT^\tnr{train}$:
($2020\text{-}11\text{-}16 \to 2022\text{-}11\text{-}17$); 
$\mcT^\tnr{test}$:
($2022\text{-}11\text{-}17 \to 2024\text{-}07\text{-}14$).
Elo uses $\tilde{s}=s/\log 10 \approx 260.6$
(FIFA $s=600$ and $\mcL(\cd;10)$).
Lower~$\mathsf{LS}$ is better.}
\label{tab:FIFA}
\setlength{\tabcolsep}{6pt}
\begin{tabular}{lcccc}
\toprule
Method & $\hat\alpha_1$ & $\hat\beta$ & $\hat\eta$ & $\mathsf{LS}$ \\
\midrule
Conventional
  & $\phantom{-}0.693$ & $0.500$ & $0.000$ & $0.998$ \\
\midrule
Simple, no HFA
  & \multirow{4}{*}{$-0.588$} & \multirow{2}{*}{$0.783$} & $0$                       & $0.904$ \\
Simple, with HFA
  &                           &                           & \multirow{3}{*}{$0.730$}  & $0.894$ \\
Optimal scaling
  &                           & $0.541$                   &                           & $0.893$ \\
On-line adaptive~\eqref{hat.beta.mini.batch.update}
  &                           & $0.596$                   &                    & $0.891$ \\
\midrule
Fully adaptive
  & $-0.301$                  & $0.502$                   & $0.826$                   & $0.891$ \\
\bottomrule
\end{tabular}
\end{table}

The results are in line with those obtained on synthetic data in Example~\ref{Ex:find_alpha_and_scale}: the conventional approach straw-mans the FIFA ranking and we can do much better using simple formulas for $\hat\balpha$  and $\hat\eta$ in \eqref{alpha.1.initialization} and \eqref{eta.0.initialization}; again, the full adaptation is not required. 

Optimal scaling unexpectedly is smaller than $\hat\beta$ from \eqref{beta.0.initialization}. It is thus instructive to inspect $\beta_t$ obtained via \eqref{hat.beta.mini.batch.update} and shown in Fig.~\ref{fig:FIFA} together with the skills of 15 teams with skills covering uniformly the range of values at the last match. 

\begin{figure}[t]
    \centering
\includegraphics[width=0.9\linewidth]{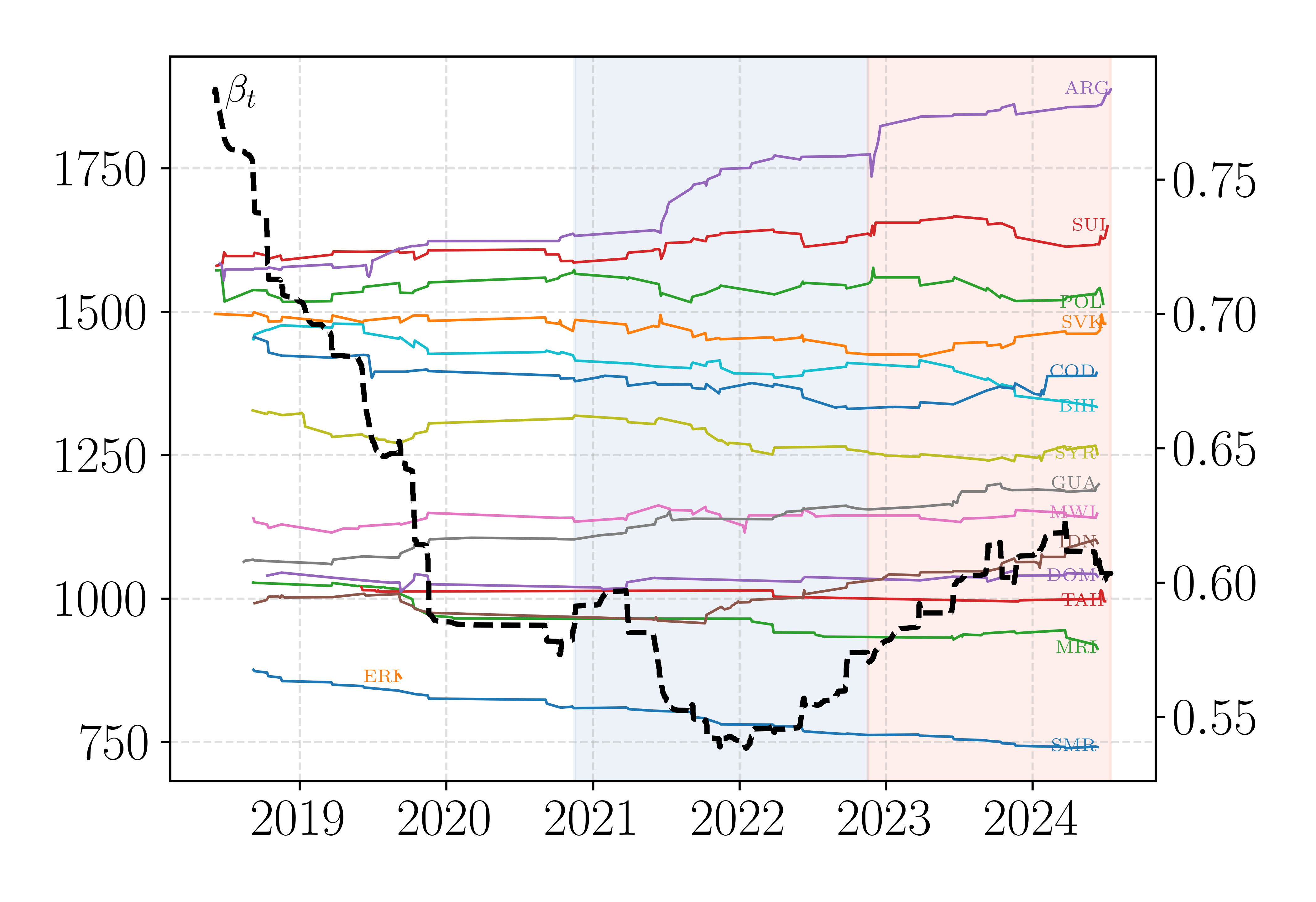}
    \caption{Skills (left axis) of the team, from the best to the worst, (thin lines and three-letters abbreviations) and $\beta_t$ (dashed thick line) estimated via \eqref{hat.beta.mini.batch.update} (right scale). The shaded regions correspond to the training/testing sets (blue/rose, respectively).}
    \label{fig:FIFA}
\end{figure}

We can visually appreciate that the skills' spread increases in time, \ie we cannot declare convergence and thus, the skills differences are too small comparing to those at the hypothetical convergence. 

This observation, already made in \cite[Sec.~5.1]{Szczecinski22a}, is more precisely assessed by $\beta_t$ ( shown also in Fig.~\ref{fig:FIFA}) calculated using \eqref{hat.beta.mini.batch.update} with $W=100$. Since the scale adjustment $\hat\beta$ is smaller than predicted by the model at convergence, rather than correcting the effect of the estimation noise, $\hat\beta_t$ amplifies too small values of $z_t$.

The lack of convergence can be well explained analyzing how many time-constants the team $m$ played on average, \ie we calculate the exponent in \eqref{convergence.in.expectation}
\begin{align}
    \Lambda_m &= \frac{N_m}{\ov{\tau}_m},
\end{align}
where $N_m$ is the number of matches played by team $m$ and $\ov\tau_m$ is given in \eqref{average.tau}. 

The cumulative distribution of $\Lambda_m$ is shown in  Fig.~\ref{fig:Lambda.FIFA} where, by 2020m, none of the teams accumulated one time constant; clearly, no convergence could be declared even in its most liberal interpretation.

Since 2022, the convergence conditions improve (note also the increase in $\beta_t$ in Fig.~\ref{fig:FIFA}), yet, by 2024, 80\% of the teams have not played enough matches to accumulate one time constant, $\ov\tau_m$, and we still had 98\% of teams with fewer matches than $2\ov\tau_m$. After six years of running the algorithm the convergence cannot be declared yet. This happens mostly because weak teams do not qualify to high-stakes matches (where $K_t$ is large) and play small number matches with slow-convergence, \eg friendlies with  $K_t\in\set{5,10}$.

\begin{figure}
    \centering
    \includegraphics[width=0.9\linewidth]{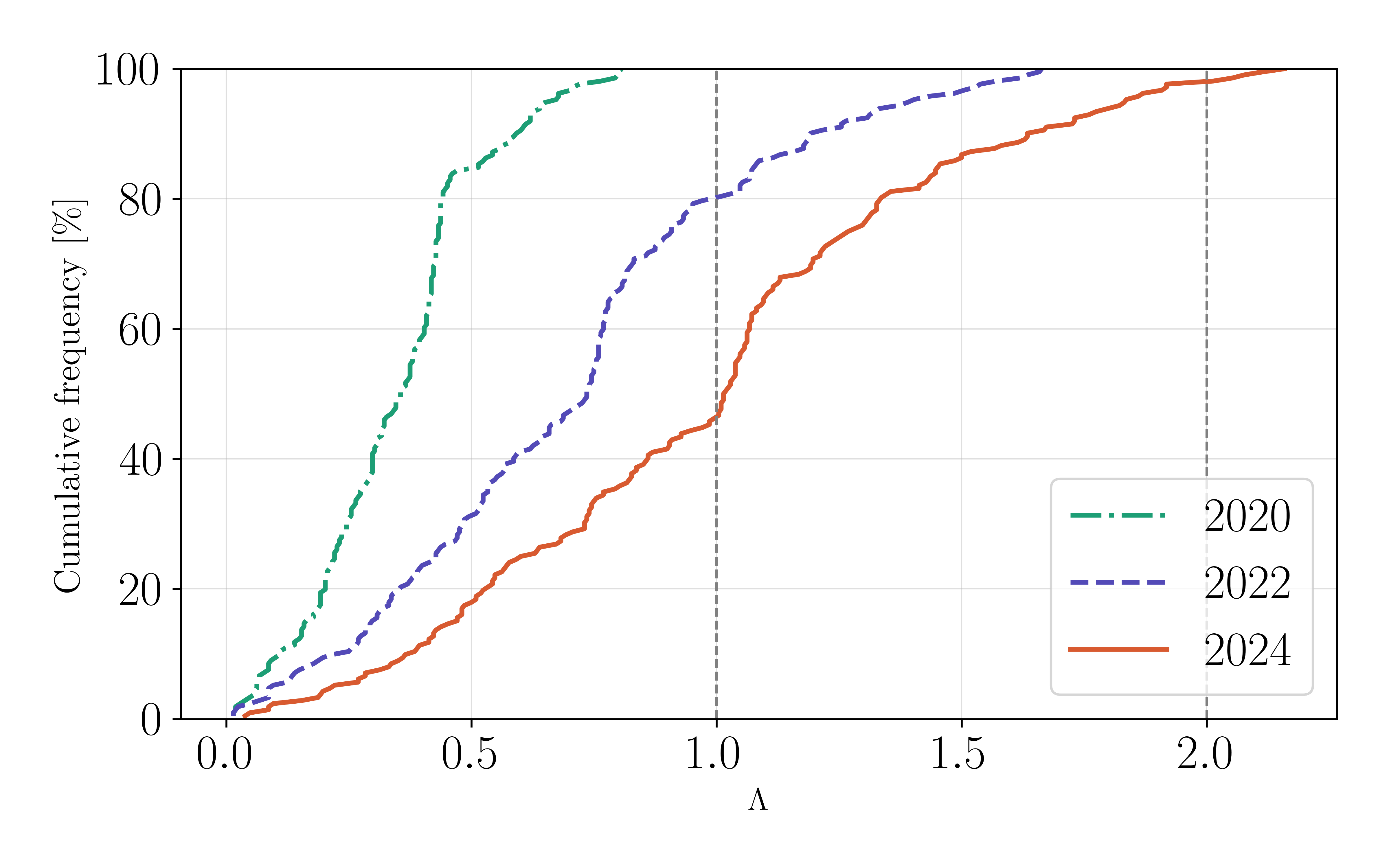}
    \caption{Percentage of international \gls{fifa} teams which have played at least $\Lambda_m=\Lambda$ [time constants] matches by the year 2020, 2022, and 2024. The vertical lines indicate number of matches equal to one and two average time constants $\ov\tau_m$, were $\Lambda\approx 2$ is required to reliably declare convergence in the mean.}
    \label{fig:Lambda.FIFA}
\end{figure}

\end{example}

\section{Conclusions}\label{Sec:Conclusions}

This work reconciles the practitioner's and statistician's perspectives on the Elo ranking algorithm around two main contributions: practitioners obtain a principled framework for performance evaluation of the algorithm, while statisticians may interpret it as approximate \gls{ml} estimation via \gls{sg} updates.

We recapitulate our finding as follows:
\paragraph{Binary outcomes.}
In the binary case (Sec.~\ref{Sec:ranking.binary}), the practitioner's update rule and the ML+SG algorithm coincide if and only if the expected score is the logistic function $\mcL(z/s)$ (Sec.~\ref{Sec:practitioner.meets.statistician}), which is thus the unique expected score endowed with a full probabilistic interpretation. The scale $s$ and step $K$ govern a fundamental trade-off: for fixed $s$, increasing $K$ reduces the convergence time constant $\tau=4s/K$ \eqref{tau.from.K} at the cost of larger asymptotic estimation variance $\ov{v}=Ks/2$. This trade-off should guide parameter choices in practice. The \gls{hfa} parameter $\eta$ does not change if we rescale the skills (for a given expected score), and may be reused in different approximations of $\mcL(z)$ if the product $\eta s$ is held constant.

\paragraph{Estimation noise forces model decoupling.}
We show that the estimation noise produced by the \gls{sg} algorithm forces a distinction between the ranking model and the prediction model even when both are otherwise identical. The effective scale used in prediction from ``noisy" estimated skills is increased $\hat{s}=s\beta_\tnr{err}>s$ \eqref{s.hat.from.s}, while the effective \gls{hfa} is attenuated, $\hat\eta<\eta$ \eqref{hat.eta.transformed}. A practitioner who reuses the nominal $s$ and $\eta$ for prediction systematically overestimates the predictive power of the model. The pseudo-model identification step~\eqref{hat.s.from.data} finds $\hat{s}=s\hat{\beta}$ and $\hat\eta$ directly from the data.

\paragraph{Multilevel outcomes.}
For $L>2$ outcomes (Sec.~\ref{Sec:Multilevel.outcomes}), the \gls{gelo} algorithm is the canonical ML+SG update for the \gls{ac} model. The Elo ranking with uniform scores ($\rho_y=y/(L-1)$) corresponds to an implicit choice of \gls{ac} parameters (Proposition~\ref{Prop:AC=Elo}). In general, the expected score $G^\tnr{AC}(z/s)$ and the logistic function $\mcL(z/\tilde{s})$ agree only approximately. For $L=3$ and draw probabilities typical in sports ($\alpha_1\lesssim 0.7$) the approximation is accurate and Elo algorithm yields reasonable estimation results which explains why the algorithm ``works" in practice.

\paragraph{Model decoupling in practice.}
The \gls{ac} model parameters corresponding exactly to the Elo ranking need not be those that best describe the data. The decoupling principle (Sec.~\ref{Sec:model.decoupling}) resolved this issue by separating estimation from prediction: the parameters $(\hat\balpha, \hat\eta, \hat\beta)$ to be used in prediction can be found from the data by treating the skills $\btheta_t$ obtained from the Elo ranking as fixed inputs~\eqref{hat.alpha.scale.batch}--\eqref{neg.log.scores.adaptive}. 

The examples on synthetic and real \gls{fifa} data show that most of the gain over the conventional, coupled models is already captured by the simple closed-form estimates \eqref{alpha.1.initialization}--\eqref{beta.0.initialization} (obtained from the match outcome frequencies), with only the scale adjustment parameter $\hat\beta$ estimated from the data.

The \gls{fifa} case further reveals that when convergence has not been reached (\ie when teams play infrequently relative to their own time constant $\ov\tau_m$) the estimated $\hat\beta$ falls \emph{below} its noise-correction baseline, which \emph{amplifies} the compressed skill differences. The on-line estimate $\beta_t$~\eqref{hat.beta.mini.batch.update} tracks this in real time and serves as a diagnostic of convergence.

The results make a case that using the same model at the same scale for both ranking and evaluation is neither necessary nor useful. The decoupling is conceptually transparent, computationally inexpensive. Examples show that it consistently improves, the predictive log-score.

\begin{appendices}
\section{Derivation of \eqref{formula.empirical.expected.score}}\label{Proof.empirical.expected.score}

From \eqref{pdf.z^*.z}, dropping the time-index $t$, by the rule of multiplication of the Gaussian distributions \cite[Sec.~8.4]{Barber12_Book}, we have
\begin{align}\label{pdf.z^*.z.formula}
    \pdf(z^*|z)
    &=\mcN(z;z^*,2\ov{v})\mcN(z^*;0,2v_\theta)/\mcN(z;0,2(\ov{v}+v_\theta))\\
    &=\mcN(z^*;z/a,2\ov{v}/a),\\
    \label{a.definition}
    a&=1+\ov{v}/v_\theta.
\end{align}

Since we use $\mcL(z/s+\eta)\approx\Phi((z+\eta s)/(s\beta_{\mcL\rightarrow\Phi}))$, see \eqref{from.Phi.to.L}, the following relationship is useful
\begin{align}\label{expectation.CDF}
    \int \Phi((x+z)/b)\mcN(x;y,q^2)\dd x
    &=\Phi\left(\frac{y+z}{\sqrt{b^2+q^2}}\right)
\end{align}
to express \eqref{ov.Pr.Y=1} as
\begin{align}
    \ov{\tnr{Pr}}\set{Y_t=1|z_t}&\approx
    \int \Phi\left(\frac{z^* + \eta s}{s\beta_{\mcL\rightarrow\Phi}}\right)\mcN\left(z^*;\frac{z}{a};2\frac{\ov{v}}{a}\right)\dd z^*\\
    &=
\Phi\left(\frac{z/a+\eta s}{\sqrt{s^2\beta_{\mcL\rightarrow\Phi}^2  +2\ov{v}/a}}\right)\\
&\approx \mcL\left(\frac{z +\eta s a}{a\sqrt{s^2+2\ov{v}/(a\beta_{\mcL\rightarrow\Phi}^2})}\right)
=\mcL\left(\frac{z}{\hat{s}}+\hat\eta\right),
\end{align}
where $\hat{s}$ is given in \eqref{s.hat.from.s} and $\hat\eta$ in \eqref{hat.eta.transformed}.

\section{Proof of Proposition~\ref{Prop:AC=Elo}}\label{Sec:Proof.Lemma.AC}

Using the binomial expansion, $(1+a)^{L-1}=\sum_{y=0}^{L-1}\nchoosek{L-1}{y}a^{y}$, $\alpha_y = \log\nchoosek{L-1}{y}$ and $\delta_y= \frac{y}{L-1}$, the expected score can be calculated as
\begin{align}
    G(z/s) & = 
    \frac{\sum_{y=0}^{L-1} \delta_y \e^{\alpha_y+\delta_y z/s}}
    {\sum_{y=0}^{L-1} \e^{\alpha_y+\delta_y z/s}}
    =
    \frac{\sum_{y=1}^{L-1} \frac{y}{L-1} \frac{(L-1)!}{y!(L-1-y)!}\e^{z y/[s(L-1)]}}
    {\sum_{y=0}^{L-1} \nchoosek{L-1}{y}\e^{z y/[s(L-1)]}}\\
    &=
    \frac{\e^{z/[s(L-1)]}\sum_{y=0}^{L-2} \nchoosek{L-2}{y}\e^{z y/[s(L-1)]}}
    {(1+\e^{z/[s(L-1)]})^{L-1}}
    =
    \frac{\e^{z/[s(L-1)]}(1+\e^{z /[s(L-1)]})^{L-2}}
    {(1+\e^{z /[s(L-1)]})^{L-1}}\\
    &=
    \frac{1}
    {1+\e^{-z/[s(L-1)]}}=\mcL\big(z/\tilde{s}\big).
\end{align}
where $\tilde{s}=s(L-1)$. This terminates the proof.

\end{appendices}

\ifdefined\ARXIV

\else
\bibliography{\CFilesBib/references_rank,\CFilesBib/IEEEabrv,\CFilesBib/references_all}
\fi

\ifdefined\JSA
\bibliographystyle{apacite}  
\fi
\ifdefined\JQAS
\bibliographystyle{\CFilesBib/IEEEabrv,\CFilesBib/DeGruyter}
\bibliographystyle{abbrvnat}  
\fi

\end{document}